\documentclass[aps, notitlepage, longbibliography, twocolumn, nofootinbib]{revtex4-2}

\usepackage[utf8]{inputenc} 
\usepackage[T1]{fontenc}    
\usepackage{hyperref}       
\usepackage{url}            
\usepackage{booktabs}       
\usepackage{amsfonts}       
\usepackage{nicefrac}       
\usepackage{microtype}      
\usepackage{lipsum}
\usepackage{amssymb, bm, amsmath}
\usepackage{mathtools}
\usepackage{xcolor, graphicx}
\usepackage{lmodern}

\newcommand{\quotes}[1]{``#1''}

\newcommand{\lastequal}{Corresponding authors. These authors contributed equally.}

\begin{document}

\title{Inspecting the interaction between HIV and the immune system\\
  through genetic turnover}
\author{Andrea Mazzolini}
\affiliation{Laboratoire de physique de l’École normale supérieure (PSL University), CNRS, Sorbonne Université, and Université Paris Cit\'e, 75005 Paris, France}
\author{Thierry Mora}
\thanks{\lastequal}
\affiliation{Laboratoire de physique de l’École normale supérieure (PSL University), CNRS, Sorbonne Université, and Université Paris Cit\'e, 75005 Paris, France}
\author{Aleksandra M Walczak}
\thanks{\lastequal}
\affiliation{Laboratoire de physique de l’École normale supérieure (PSL University), CNRS, Sorbonne Université, and Université Paris Cit\'e, 75005 Paris, France}

\begin{abstract}
Chronic infections of the human immunodeficiency virus (HIV) create a very complex co-evolutionary process, where the virus tries to escape the continuously adapting host immune system. 
Quantitative details of this process are largely unknown and could help in disease treatment and vaccine development.
Here we study a longitudinal dataset of ten HIV-infected people, where both the B-cell receptors and the virus are deeply sequenced.
We focus on simple measures of turnover, which quantify how much the composition of the viral strains and the immune repertoire change between time points.
At the single-patient level, the viral-host turnover rates do not show any statistically significant correlation, however they correlate if the information is aggregated across patients.
In particular, we identify an anti-correlation: large changes in the viral pool composition come with small changes in the B-cell receptor repertoire.
This result seems to contradict the naive expectation that when the virus mutates quickly, the immune repertoire  needs to change to keep up.
However, we show that the observed anti-correlation naturally emerges and can be understood in terms of simple population-genetics models.
\end{abstract}

\maketitle

\section{Introduction}

The adaptive immune system has been shaped by evolution to provide an effective response against a practically infinite reservoir of pathogens.
During an infection, B-cells undergo affinity maturation in lymph-node germinal centers \cite{maclennan1994germinal, allen2007germinal}.
This mechanism is a Darwinian evolutionary process, where B-cell receptors are subject to somatic hypermutations \cite{campbell2013properties} and are selected depending on their ability to recognize an external pathogen.  
This increases the affinity of naive B-cells against the pathogen up to 10-100 factors \cite{victora2012germinal, shlomchik2012germinal, mesin2016germinal}, generating memory and plasma B-cells.

During chronic infections of the human immunodeficiency virus (HIV), the immune response is dominated by the action of antibody-secreting plasma B-cells \cite{mcmichael2010immune}.
However, most of the time, this machinery is not enough to control or clear the virus.
The reason can be identified in the extremely rapid evolution of the virus escaping  immune adaptation \cite{fauci2003hiv, richman2003rapid, moore2009limited} and the fact that regions of the viral structure sensitive to B cell targeting are made inaccessible \cite{kwong2002hiv, lyumkis2013cryo}.
Nevertheless, an effective immune response can instead occur naturally in 10-20\% of the patients, and it is related to the emergence of broadly neutralizing antibodies \cite{simek2009human, liao2013co, mccoy2017identification}.
This promising discovery is the basis of the search for an HIV vaccine \cite{walker2009broad, walker2011broad, kwong2013broadly, klein2013antibodies}.

The current picture is that of the two populations of HIV and B-cell repertoire undergoing rapid and complex antagonistic coevolution.
A lot of effort has been put in quantitatively understanding evolutionary properties of these two populations, which are usually considered separately.
For example, previous work has studied the dynamics of HIV variants escaping the immune system \cite{fischer2010transmission, henn2012whole, barton2016relative}, or diversity patterns and linkage equilibrium properties of the virus \cite{zanini2015population}.
On the immune system side, a large body of work has been dedicated to studying the immune response to HIV and the emergence of broadly neutralizing antibodies \cite{mouquet2014antibody, victora2018primary, kreer2022determining}, as well as to characterize lineage evolution during the affinity maturation process, using high-throughput sequencing of B-cell repertoire  \cite{hoehn2015dynamics, nourmohammad2019fierce}.
Much less work focuses on the coupled evolutionary dynamics of the two populations. Co-evolutionary work has typically been theoretical, with a general focus on BnAb generation~\cite{wang2015manipulating, nourmohammad2016host, molari2020quantitative}. 
The datasets used for the data based studies contain sequences of either HIV or  immune repertoires.
Given the interacting nature of this coevolutionary process, genetic data of both the populations evolving in time would provide valuable information.

To our knowledge, only one dataset of this kind has been made public, where a portion of the HIV envelope gene and B-cell receptors have been both deeply sequenced in time for different patients \cite{strauli2019genetic}.
We base our analysis on this dataset and  define simple macroscopical observables for the evolution of the two populations, which quantify genetic turnover and selective pressure.
We find that a few of those measures display temporal correlations between the viral population composition and the immune system, showing that the co-evolutionary \quotes{arms-race} leaves traces at the whole-population level.
To make sense of these correlations, the second part of the manuscript introduces population-genetics models with different levels of complexity.
Through numerical and analytical analysis, these models show that the observed statistical patterns can  emerge for a biologically reasonable set of parameters.
These theoretical models can help to build intuition about how and under which conditions these correlation arise.

\section{Methods}
\subsection{Longitudinal data for HIV and the immune repertoire}

Our study is based on the dataset described in \cite{strauli2019genetic}.
The samples originate from 10 HIV-infected male participants.
For each individual we have 10 to 20 longitudinal samples, taken before administration of antiretroviral therapy.
For most of the time points the viral genetic composition and the immune repertoire are tracked in parallel, see Fig.~\ref{fig:SM_data_descr} and section \ref{sec:SM_dataset_descr} for more information. 
Specifically, on the viral side, the C2-V3 region of the \textit{env} gene is deeply sequenced. This gene is known to be a potential target of the antibody repertoire \cite{hatada2010human, ringe2012unique}.
On the immune repertoire side, the samples correspond to the deep sequencing of the variable region of the immunoglobulin (Ig) heavy chain locus. 

We wrote a pipeline to download the dataset, assemble the HIV sequences and the immunoglobulin heavy chain clonotypes.
The description of our pipeline is in section \ref{sec:SM_dataset_process} of the SM. The public repository \url{https://github.com/statbiophys/HIV_coevo.git} contains all the scripts and the instructions necessary for running the pipeline and reproducing our results. 
The obtained dataset is composed of HIV samples having an average number of 1000 unique sequences and 25000 total sequences.
For the B-cell receptors there are 13000 unique clonotypes and 210000 total clonotypes.
The Ig clonotypes have been clustered into lineages as described in section \ref{sec:SM_lineages} of SM. 

\subsection{Simple measures for quantifying evolutionary properties of HIV and the immune repertoire}

We ask whether the evolutionary dynamics of HIV are temporally correlated with those of the immune repertoire.
To this end, we define and explore a few simple measures characterizing the evolution of the two populations.

\begin{figure}
\centering
\includegraphics[width=0.45\textwidth]{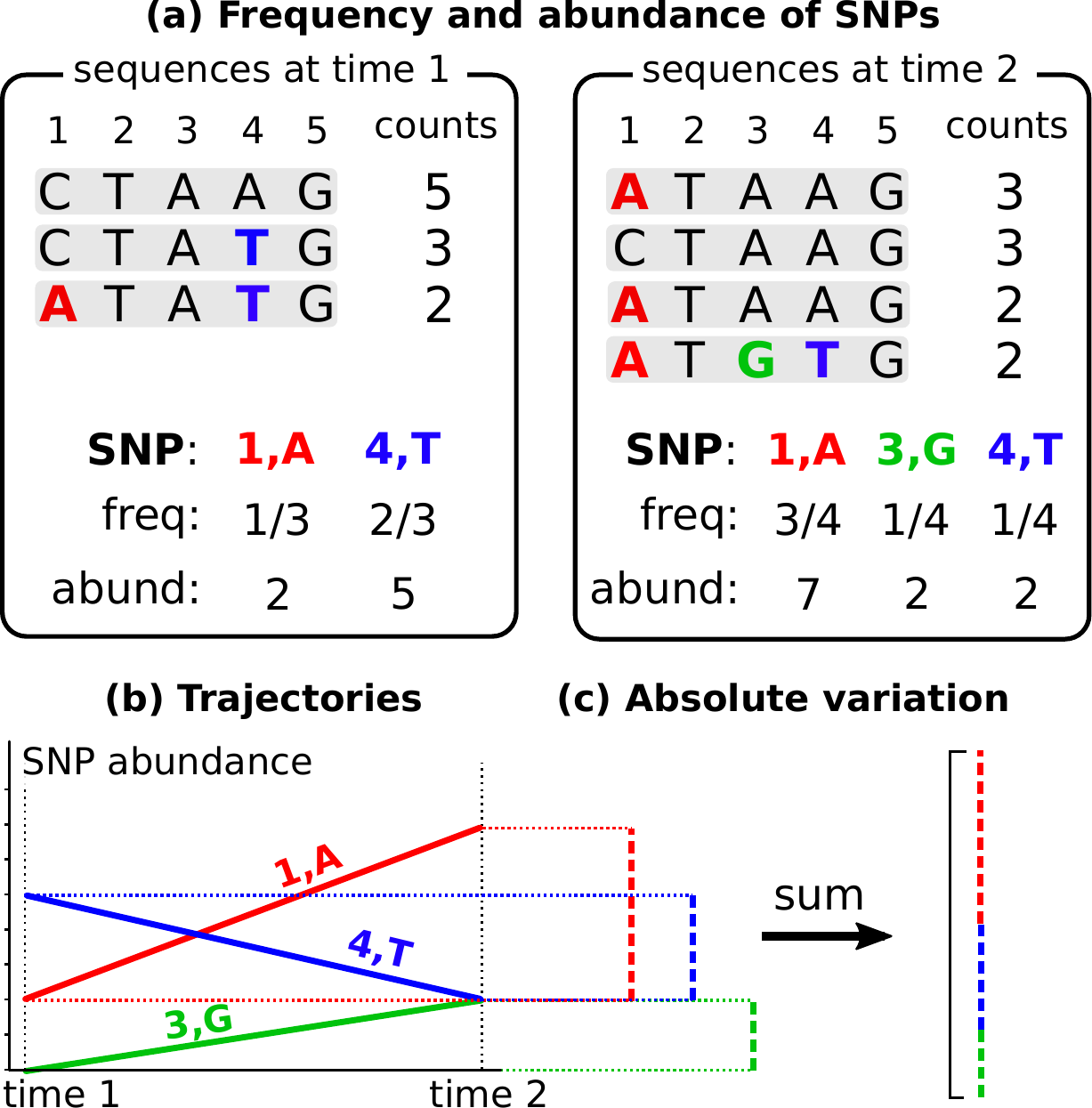}
\caption{(a) Definition of SNP frequency and SNP abundance in a toy example.
(b) Trajectories of SNP abundances. The values correspond to the example in (a).
(c) Absolute-variation computation of the trajectories. The dashed vertical lines are the different contributions to the variation given by each trajectory.
They are summed together leading to the final absolute variation.}
\label{fig:turnover_sketch}
\end{figure}

On the viral side, most of these measures are based on how Single Nucleotide Polymorphisms (SNP) of the viral sequences change in time. Details about how the SNPs are tracked in data are discussed in section \ref{sec:SM_SNPs} of SM. 
As sketched in Fig.~\ref{fig:turnover_sketch}a, for a given position in the sequences and a given nucleotide, we can count its frequency and its abundance.
The frequency is the number of unique sequences in which the nucleotide appears divided by the total number of sequences.
The abundance is the sum of the sequence counts of all the sequences in which the letter appears.
These numbers can be computed for a given SNP, $i$ (e.g. $i=(1,A)$ for an A at position 1) at different time points to create a trajectory $x_i(t)$ (Fig.~ \ref{fig:turnover_sketch}b).

To quantify how much the genetic composition of a set of sequences varies between two timepoints $t_1$ and $t_2$, we add together the absolute difference  of all the single SNP trajectories, Fig.~\ref{fig:turnover_sketch}c. 
We call this quantity \textit{absolute variation}:
\begin{equation}
\text{absolute variation}(t_1, t_2) = \sum_i |x_i(t_2) - x_i(t_2)| .
\label{eq:abs_var}
\end{equation}
The choice of the absolute value makes no distinction between an increase or a decrease of the trajectory of the same amount.
This is because we are only interested in the magnitude of the change and not in its sign.

The absolute variation can be applied to the HIV sequences of a patient across time points, both for SNP frequencies and SNP abundances.
This defines the first two entries of Table \ref{tab:hiv_measures}: \textit{HIV turnover fr/ab}.
Since these measures compute how much new mutations spread in the population from one time point to the other, they can be interpreted as the genetic turnover.

In a similar way, these measures can also be applied to the SNPs of Ig repertoire (entries \textit{Ig turnover fr/ab} of Table \ref{tab:abr_measures}). 
However, in that case, SNPs have to be computed at the level of lineages, obtaining trajectories for SNP $i$ in lineage $l$, $x_{i,l}(t)$. Note that for \textit{Ig turnover fr}, the frequency is normalized by the number of sequences in the lineage, so that $\sum_{i \in l} x_{i,l}(t) = 1$.
The absolute turnover then sums across all SNPs and lineages.

To quantify the fact that B cell lineages themselves are subject to turnover, and rise and fall in time, we introduce a measure that computes the absolute variation of lineage sizes.
The size of a lineage is the sum of all its sequence abundances or frequencies (here normalized by their total counts in the sample, so that  $\sum_{i} x_{i,l}(t) = 1$). Applying the formula for absolute turnover, Eq.~\ref{eq:abs_var}, to these quantities defines the  \textit{Ig lineage turnover fr/ab} measures in Table \ref{tab:abr_measures}.
We define two additional measures derived from the Ig lineage turnover abundances:  one that includes only the top 10\% changes in abundances in the computation of the absolute variation (\textit{Ig lineage large turnover}), and one with only the bottom 50\% (\textit{Ig lineage small turnover}). 
In general we expect that smaller variations are more subject to noise and, in turn, carry less signal, which is verified by our correlation analysis.

Finally, we define an estimate of the selective pressure on a population, closely related with the well-known measure of adaptation dN/dS 
\cite{nei1986simple, yang2000statistical}.
To this end, we compute separately the absolute variation of the non-synonymous SNPs and synonymous ones and we take their ratio:
\begin{equation}
\frac{d N}{d S}(t_1, t_2) = k \frac{\sum_{i\in \text{non-syn}} |x_i(t_2) - x_i(t_2)|}{\sum_{j\in \text{syn}} |x_j(t_2) - x_j(t_2)|}.
\label{eq:dn/ds}
\end{equation}
The coefficient $k$ is the ratio of the probabilities of randomly generating synonymous and non-synonymous mutations from the reference sequence. 
This fixes the ratio to $1$ in the case of uniform random mutations and a small mutation rate (see  section \ref{sec:SM_SNPs} of SM).
We computed this measure only for HIV and not for the immune system.
The reason is that most of the lineages are small and it can happen that their synonymous variation is zero, $dS = 0$, leading to several undefined values.

\begin{table}
\centering
\caption{List of HIV evolutionary measures}
\begin{tabular}{c|c}
\textbf{Short name} & \textbf{Definition} \\
\hline\hline
HIV turnover fr & Abs variation of SNPs  frequencies \\
\hline
HIV turnover ab & Abs variation of SNPs abundances \\
\hline
HIV dN/dS fr & Ratio of the non-syn and syn abs \\
 & variations for SNPs frequencies \\
\hline
HIV dN/dS ab & Ratio of the non-syn and syn abs \\
 & variations for SNPs abundances \\
\end{tabular}
\label{tab:hiv_measures}
\end{table}

\begin{table}
\centering
\caption{List of Ig-repertoire evolutionary measures}
\begin{tabular}{c|c}
\textbf{Short name} & \textbf{Definition} \\
\hline\hline
Ig turnover fr & Abs variation of SNPs  frequencies \\
\hline
Ig turnover ab & Abs variation of SNPs abundances \\
\hline
Ig lineage turnover fr & Abs variation of lineage frequencies \\
 \hline
Ig lineage turnover ab & Abs variation of lineage abundances \\
\hline
Ig lineage large & The largest 10\% abs variation \\
turnover & of lineage abundances \\
\hline
Ig lineage small & The smallest 50\% abs variations \\
turnover & of lineage abundances \\
\end{tabular}
\label{tab:abr_measures}
\end{table}

\section{Results}

\subsection{A single-patient analysis does not show any interaction}\label{single_patient}
We will show that the co-evolutionary interaction between HIV and the immune system leaves a trace in the dynamics of two populations.
More precisely, some of the evolutionary measures defined for HIV, Table \ref{tab:hiv_measures},  significantly temporally correlate with properties of the immune repertoire, Table \ref{tab:abr_measures}.

However, this signal is almost not visible at the single patient level.
For example,  in Fig.~\ref{fig:SM_corr_single_pat} of the SM, we compute  the Spearman correlation between the \textit{HIV turnover ab} and the \textit{Ig lineage turnover ab} trajectories in the 9 patients.
Eight out of 9 patients show an anti-correlation, but the signal is weak, with only 3 out of  8 showing a p-value below 0.1, and only 1 (patient 4) showing a significant correlation with $p=0.004$ (which comes very close to the significance threshold of 0.05 after the Bonferroni correction).
Nevertheless, this kind of analysis is not catching the fact that all these weak correlations point towards the same direction.
Below we propose a statistical measure which shows that the correlation between these pairs of measures deviates significantly from a null model where no correlations exist.

\subsection{Statistical procedure for combining temporal correlations across patients}
\label{sec:ks_stat_test}

\begin{figure*}
\centering
\includegraphics[width=0.9\textwidth]{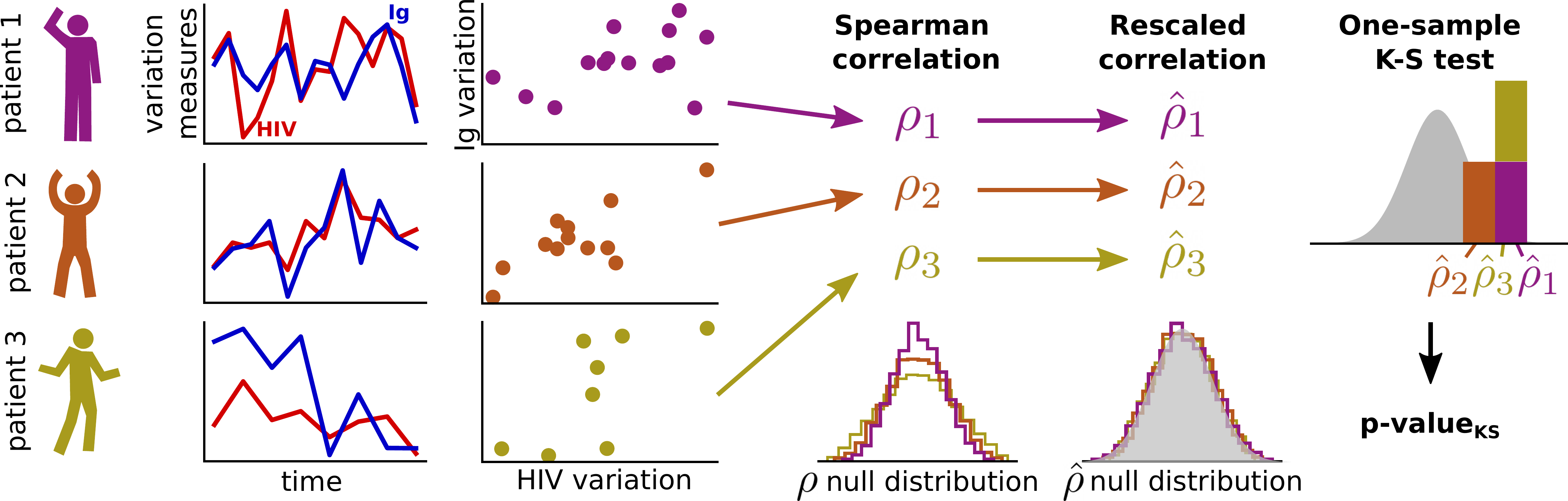}
\caption{Cartoon of the statistical procedure to integrate correlations across different patients (not real data). 
Given two measures of variation (one for the HIV and one for the Ig repertoire) computed in all the patients, the Spearman correlations between the trajectories are computed and then re-scaled in such a way that their null distributions collapse across patients.
Finally, a one-sample Kolmogorov-Smirnov test verifies if these re-scaled coefficients deviate significantly from the null distribution of correlations.}
\label{fig:ks_stat_test}
\end{figure*}

As we saw in section~\ref{single_patient} and Fig.~\ref{fig:SM_corr_single_pat} of the SM, the correlations between the considered measures  were negative for all patients but with a non-significant p-value, which led us to conclude that there is no significant correlation.
However, it is very unlikely that, if the two measures are really independent, they generate a coherent positive (or negative) correlation across all patients.
In the following we describe a procedure whose aim is to quantify this observation of co-variation and \quotes{integrate} the information contained in coherent correlations. 

We graphically illustrate the statistical integration procedure in the cartoon of Fig.~\ref{fig:ks_stat_test} using an artificial dataset.  We calculate the Spearman correlation, $\rho$, for a pair of variation measures in three patients.
In the first step we re-scale the correlation coefficients $\rho$ between the viral and Ig measures from the different patients in a way that they become comparable.
Specifically we define a new variable, $\hat{\rho}$, whose null distribution is the same across all patients.
The null model is built by reshuffling the $n$ time points of the two measures in each patient. The distribution of Spearman correlation coefficients obtained in this way follos approximately a Gaussian law centered at zero with a standard deviation dependent on $n$, $\sigma_n$.
Therefore the choice $\hat{\rho} = \rho/\sigma_n$ creates a set of coefficients whose null counterparts come from the same normal distribution, which does not depend on $n$ anymore.
The second step is to test if this set of re-scaled coefficients deviates significantly from the null uncorrelated scenario.
To do so, we perfom a one-sample Kolmogorov-Smirnov test against this null normal distribution. The obtained p-value quantifies this significance. 

\subsection{HIV turnover and Ig lineages turnover are significantly anti-correlated}\label{corr1}

\begin{figure*}
	\centering
	\includegraphics[width=\textwidth]{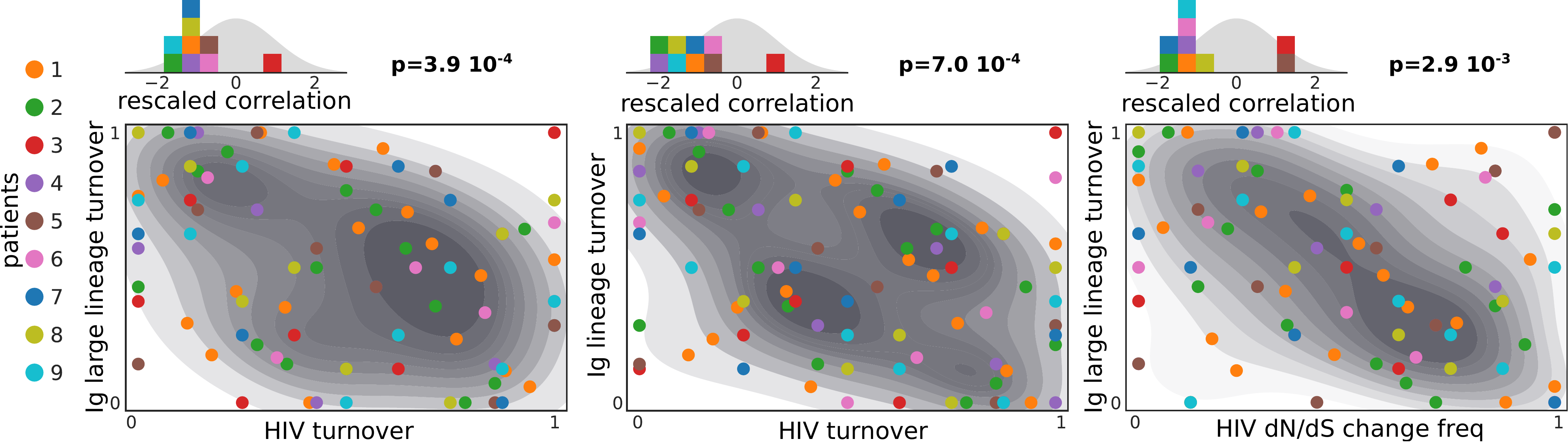}
	\caption{The three significant pairs of (anti)correlated measures. The scatter plots show the rank (normalized between 0 and 1) of all the values of the two variation measures in all the patients (differentiated by the color of the dots).
		The shaded areas below are the densities of these points computed with a kernel density estimate.
		Above each scatter plot, the histogram of the re-scaled correlations is shown, compared with the null distribution (shaded grey area).
		This is the same plot as the left-most one of \ref{fig:ks_stat_test}.
		They all show an average negative re-scaled correlation.
		The Kolmogorov Smirnov p-values are displayed next to the histograms.}
	\label{fig:ks_corr}
\end{figure*}

To combine information across patients, we employ the statistical procedure explained in section \ref{sec:ks_stat_test}. 
For each pair of the HIV-Ig variation measures, we compute a Kolmogorov-Smirnov p-value on the distribution of re-scaled correlation coefficients. 
The 24 p-values are shown in Fig.~\ref{fig:SM_ks_p_all}.
We correct for multiple testing through a Benjamini–Hochberg test at a false discovery rate of $0.05$.
This selects three significant pairs of measures, which are shown in Fig.~\ref{fig:ks_corr}. 
The first two significant pairs are the Ig lineage turnover and the HIV turnover.
We find a correlation when all the lineage turnovers are considered (middle plot, \textit{Ig lineage turnover ab}), or when only the largest lineage variations are taken (left plot, \textit{Ig lineage large turnover}).
The signal disappears if one considers only small variations, \textit{Ig lineage small turnover}, Fig.~\ref{fig:SM_ks_p_all}, a measure that is probably more susceptible to noise.
It is interesting to see that the correlation is negative, as can be seen from the small histograms on the top of Fig.~\ref{fig:ks_corr} and from the grey density of all the points of the patients put together in the scatter plots.
The third significant pair correlates lineage turnover versus the dN/dS measure of HIV.
This group of re-scaled correlations points again towards an anti-correlation.

The fact that the correlation between turnovers is negative seems counter-intuitive.
It means that if the composition of the viral population is changing quickly, the immune repertoire abundances are not changing much, while the immune repertoire changes its composition faster when the viral population is varying more slowly.

We performed additional tests to verify that the observed signals are not caused by spurious effects of the data.
A possible confounding factor is sequencing depth: if the sizes of the HIV and Ig samples are correlated for some reason, and one of the considered measures is, in turn, correlated with size, a spurious correlation can appear.
However, as shown by Fig.~\ref{fig:SM_ks_p_all}, the number of HIV sequences does not correlate with the number of Ig clonotypes as well as with any of the Ig measures (and similarly for the Ig number of sequences).
Another spurious correlation can be generated by the fact that the time points are not homogeneously distributed (Fig.~\ref{fig:SM_data_descr}), and the considered variation measures are dependent on the length of the time windows.
However, we find that the size of time windows (\textit{Delta time} in Fig.~\ref{fig:SM_ks_p_all}) does not show a correlation with any of the other measures.

Conversely, the Benjamini–Hochberg procedure we use to account for multiple testing makes the conservative assumption that the different tests we tried are independent of each other. 
But it is likely that many of the defined quantities are strongly correlated, e.g. \textit{Ig lineage turnover ab} and \textit{Ig lineage large turnover}, making the effective number of tests smaller than actually used in the procedure, leading us to overestimate the corrected p-values.
Section \ref{sec:SM_null_corr} of SM discusses this problem using a different null model which generates trajectories that reproduce the internal correlations present within the HIV measures and within the Ig measures.
This test confirms that the observed patterns are strongly unlikely to be generated by the refined null model.

\subsection{The HIV affects the future state of the immune system}

\begin{figure}
\centering
\includegraphics[width=0.45\textwidth]{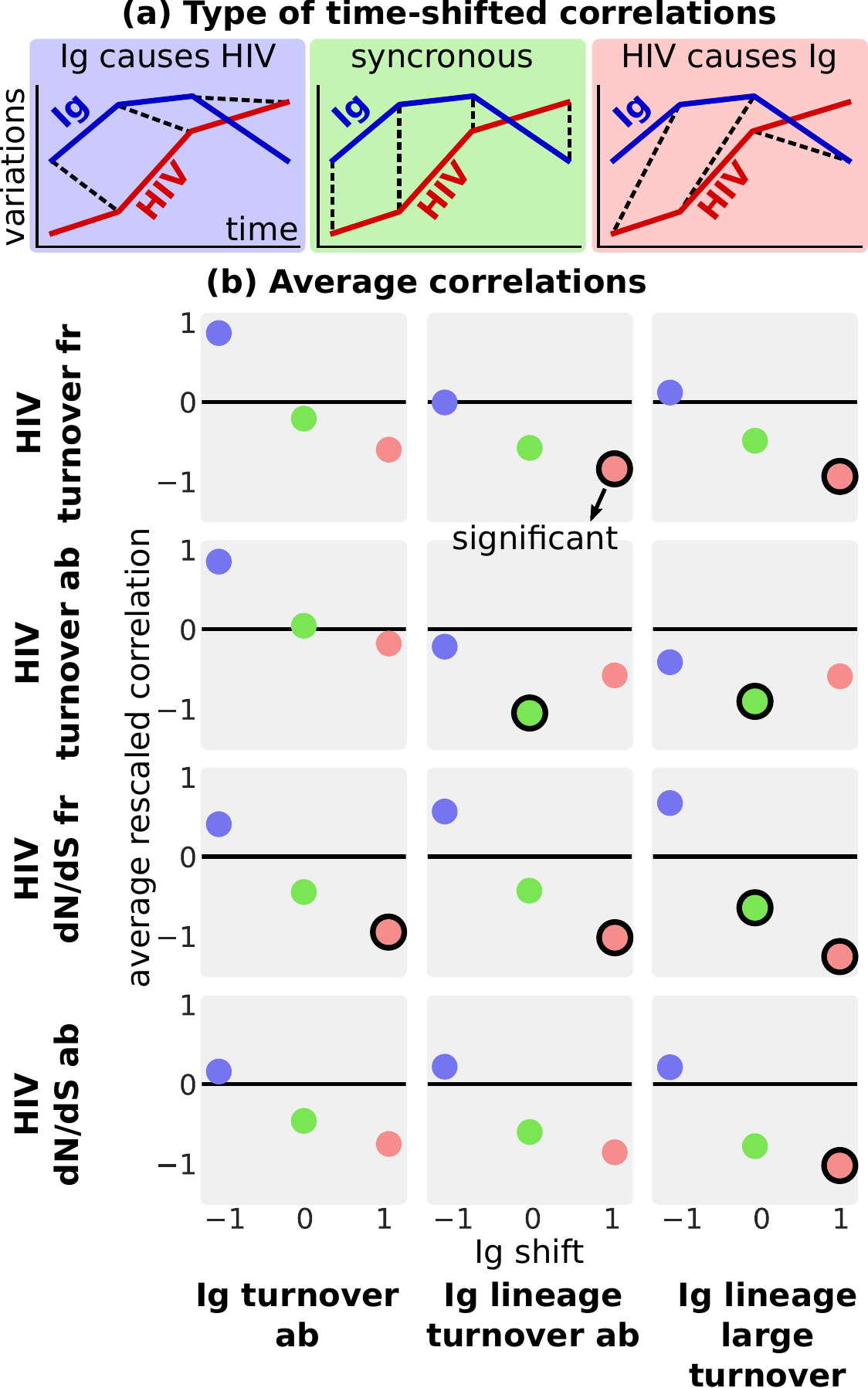}
\caption{Testing correlations across time, using temporal shifts of the trajectories.
(a) Cartoon of the three considered scenarios: HIV population change follows repertoire change (Ig shift -1), synchronous change (Ig shift 0), repertoire change follows HIV population change (Ig shift +1).
(b) For each pair of HIV-Ig measures, the average re-scaled correlation each Ig time shift.
The dots with a black border are significant according to the Benjamini–Hochberg test with a false discovery rate of $0.05$.
The pairs that do not show any significant correlation are shown in Fig.~\ref{fig:SM_ks_p_all_shift}.}
\label{fig:ks_time_shift}
\end{figure}

We repeat the analysis of section~\ref{corr1} but with a temporal shift to the trajectories. 
In particular, we compute the correlations between the points of the Ig trajectories against the points of the HIV ones, but one step forward in time (Ig time shift +1, blue panel of Fig.~\ref{fig:ks_time_shift}) or one step back (Ig time shift -1, red panel of Fig.~\ref{fig:ks_time_shift}).
The central green panel corresponds to the case considered in the section~\ref{corr1} (no time shift).

The statistical procedure discussed in section \ref{sec:ks_stat_test} is then applied in all these cases, leading to a distribution of re-scaled correlations whose average is plotted in the lower panels of Fig.~\ref{fig:ks_time_shift}.
The significant pairs (Benjamini–Hochberg test with 5\% false discovery rate) are highlighted with a black border.
The p-values for all the pairs are displayed in Fig.~\ref{fig:SM_ks_p_all_shift}.
In addition to the previously observed significant correlations with no time shift, significance is achieved by 6 pairs, in which changes in HIV precedes changes of the immune system at the future time step (red dots).
In other words, a larger turnover of HIV is followed by a smaller turnover of the Ig lineages at a later time.
The average time of this delay, i.e. the average time distance between consecutive points of the dataset, is around 225 days.
No pair shows a significant correlation in the opposite case, suggesting that the immune system does not immediately affect the future dynamics of the HIV population.

\subsection{Turnover correlations can be reproduced by a population genetics model}
\label{sec:pop_gen_model}

Assuming a co-evolutionary arms race, one could have expected a positive correlation between HIV and Ig repertoire turnovers, in contradiction with the anti-correlation observed in Figs.~\ref{fig:ks_corr} and ~\ref{fig:ks_time_shift}. When one population is changing a lot, the other population is expected to change as well to keep up, leading to a positive correlation. In the following we show using population-genetics models that this intuitive argument is not generally correct, and that anti-correlations can emerge for a reasonable range of parameters.

We investigate the  emergence of correlation patterns in genetic turnovers within a model used before in the context of HIV and  immune system coevolution \cite{nourmohammad2016host}, with a few minor modifications.
The two populations are characterized by binary strings of length $L$.
The genotype of a virus $v$ is denoted by $v = (\sigma_i^v, \sigma_2^v, \ldots, \sigma_L^v)$, where $\sigma_i =1$ or $-1$. 
Similarly, a clonotype of a B-cell receptor is characterized by a binary string $r = (\sigma_i^r, \sigma_2^r, \ldots, \sigma_L^r)$.
We consider populations of fixed size, $N_V$ and $N_R$. The fraction of a given virus strain or Ig clone are denoted by $x_v$, $x_r$.
The ability of a B cell to recognize a viral strain and expand depends on how much its Ig string is similar to that of the virus.
At the same time, a virus strain proliferates more easily if it is different from the receptor sequences.
We define the affinity between $v$ and $r$ as follows:
\begin{equation}
E_{v,r} = \sum_{i=1}^L \sigma_i^v \sigma_i^r .
\label{eq:affinity}
\end{equation}
Affinity enters the definition of fitness for the two populations, which is proportional to its average over the whole adversary population:
\begin{equation}
f_R = \frac{s_R}{2}\sum_v x_v E_{r,v}, \hspace{0.5cm} f_V = - \frac{s_V}{2}\sum_r x_r E_{r,v},
\label{eq:linear_fitness}
\end{equation}
where the two selection coefficients $s_R, s_V > 0$ control the strength of selection.

This antagonistic co-evolutionary dynamics is simulated using a Wright-Fisher model, whose details are described in section \ref{sec:SM_WF} of SM.
Briefly, in one generation, each genotype $i$ generates a binomially distributed number of offspring, with a probability proportional to the exponential fitness (given by Eq.~\ref{eq:linear_fitness} for Ig, or its negative value for HIV) times the genotype frequency, $\exp(f_i) x_i$. 
The total number of individuals in the population is fixed and imposed via multinomial sampling of the population from one generation to the next. 
After each reproduction step, there is a probability that a given site switches sign, leading to a mutated offspring.
We call the mutation rates per site per generation $\mu_V$ and $\mu_R$.

We performed simulations using values for the population size and the mutation rate  estimated from data.
The effective population size of viral population and number of B-cells in germinal centers is approximately $N_R \sim 10^3-10^4$, $N_V \sim 10^2-10^{3}$ \cite{tas2016visualizing, molari2020quantitative, lemey2006hiv}.
The mutation rates are estimated from neutral sequence diversity as $\mu_R \sim \mu_V \sim 10^{-5}-10^{-4}$ over a length of $L_R \sim 200$ for the receptor variable region, and $L_V \sim 1000$ for the \textit{env} protein of HIV. 
This leads to $1-10$ mutations per generation in both populations \cite{nourmohammad2016host, elhanati2015inferring, zanini2015population} 

As done in \cite{nourmohammad2016host}, in our model we need to set the same $L$ for both the population, choosing $L=50$ amino-acids, which means a gene of a similar length of the B cell receptor variable region.
The other parameters are chosen in a way to keep the total number of mutations per generation equal to $1$ in both the populations: $N_R = N_V = 10^{3}$, $\mu_R = \mu_V = 10^{-3}/L$.

\begin{figure}
\centering
\includegraphics[width=0.48\textwidth]{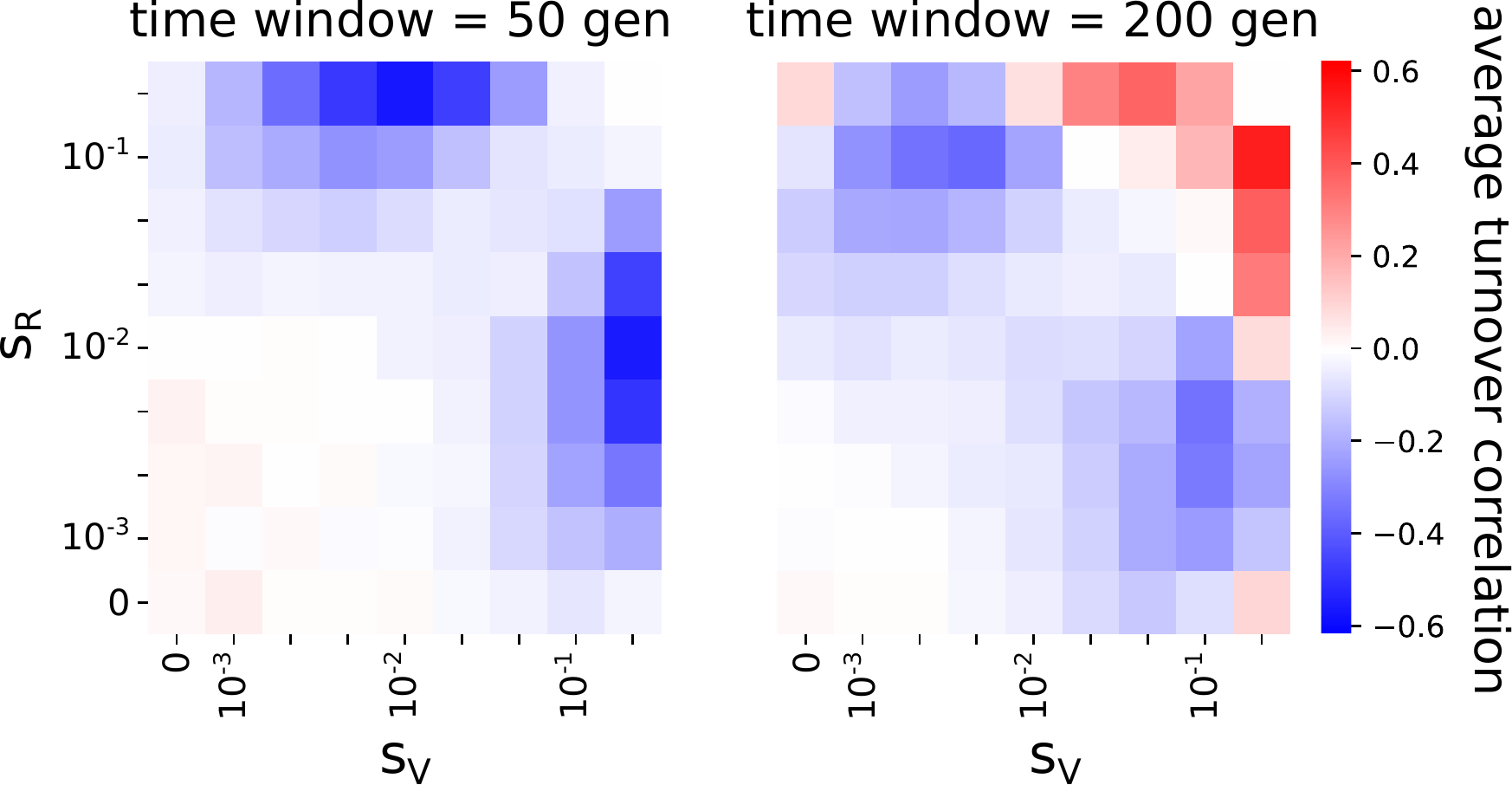}
\caption{Color map of the average re-scaled correlation of the genotype absolute variation in the co-evolving populations of binary strings.
The parameters are $N_R = N_V = 10^{3}$, $\mu_R = \mu_V = 10^{-3}/L$, $L=50$.
For each set of parameters an ensemble of $10,000$ realizations is generated.
The absolute variation of genotype frequencies is computed by considering $20$ bins of times separated by $50$ (left) or $200$ (right) generations.
Given the two trajectories of absolute variations, the Spearman correlation is computed, re-scaled and averaged across realizations. 
Details of the simulations are given in section \ref{sec:SM_WF} of SM.
}
\label{fig:bin_str_corr}
\end{figure}

Fig.~\ref{fig:bin_str_corr} shows the correlation between turnovers generated by the simulation.
The genotype frequencies of the two populations are sampled every $50$ or $200$ generations. 
The real generation time of B cells or HIV can be of order of one day \cite{molari2020quantitative}, meaning that we sample our simulations every few months, similarly to what was done in the experimental data.
We compute the absolute variation by considering  $20$ consecutive samples, using Eq.~\ref{eq:abs_var}, where $x$ corresponds to the genotype frequency.
We then correlate the  two obtained trajectories of absolute variations for the two populations and plot 
 the average of those re-scaled correlations over repeated realizations of the simulation.
The behavior is complex, depending on the selection coefficients and the interval of time chosen for the sampling.
However, Fig.~\ref{fig:bin_str_corr} shows that the region of parameter for which the correlation is negative is wide, consistent with empirical observations.

\subsection{Minimal population genetics models can reproduce the observed signal}
\label{sec:telegraph}

To further gain intuition about the observed turnover correlations, we make a strong simplification of neglecting the mutational background~\cite{neher2013genetic} and consider the antagonistic coevolution of two populations each with only one locus. This corresponds to the previous model with $L=1$.

\begin{figure}
	\centering
	\includegraphics[width=0.48\textwidth]{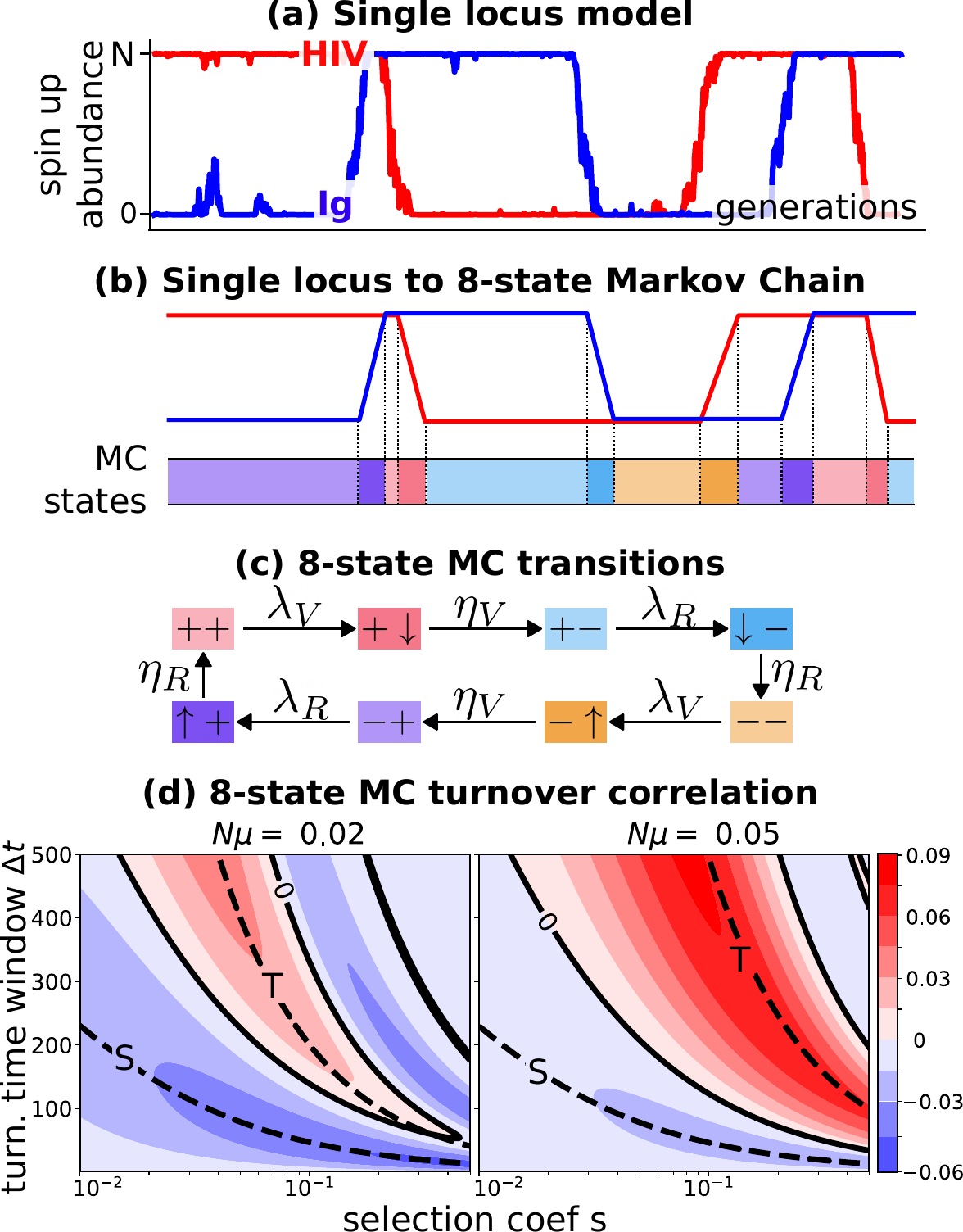}
	\caption{(a) Number of up states of two populations of antagonistically interacting loci (model of Sec.~\ref{sec:pop_gen_model} with $L=1$).
		The parameters are  $N=N_R = N_V = 10^{3}$, $\mu=\mu_R = \mu_V = 10^{-3}/50$, $s=s_R=s_V=0.005$.
		(b) Schematized trajectories where the dynamics is composed of periods of fixed populations, interleaved with periods of switches to the next fixation.
		(c) Equivalent 8-state Markov Chain whose states are represented with different colors shown in (b).
		 (d) Analytical solution for the turnover correlation (Eq.~\ref{eq:SM_corr_toy} of SM) for two populations with identical size, mutation, and selection parameters, as a function of the selection coefficient $s$ and of the time window for turnover computation $\Delta t$.
		Black lines indicate zero correlation.
		The dashed lines are the establishment time, $T$, and the switch time $S$ as a function of $s$.
		The two different plots refer to two choices of $\mu N$, with $N=10^3$.}
	\label{fig:telegraph}
\end{figure}

Using parameters corresponding to the full model, within the simple model each population alternates between states of almost complete fixation and  switching (Fig.~\ref{fig:telegraph}a), characteristic of the successional-mutation or selective sweep regime  \cite{desai2007beneficial, ebert2020host}. 
Within this regime, the simple model estimates the time for finding and establishing a beneficial mutation as $T \sim 1/(s N \mu)$. This time can be applied to HIV when its population is fixed and the spin sign is concordant with that of the immune system. For the immune system, it applies when its sign opposite that of HIV. In both cases mutations are beneficial with a selective advantage $s$. The time for a beneficial mutation to expand in the whole population scales as $S \sim \log (s N) / s$~\cite{desai2007beneficial}.
The selective sweep regime, $S \ll T$,  means that each sweep  finishes before the sweep of the other population starts, which corresponds to the limit of weak mutations $N \mu \ll 1$.

This simply model maps the trajectories in Fig.~\ref{fig:telegraph}a onto an effective 8-state discrete Markov-Chain (Fig.~\ref{fig:telegraph}b,c).
Each of the two populations can either be at a fixed or sweep state, which combined with the four possible all $+$ and all down configurations of the two populations lead to 8 possible states.
Transitions to the sweep states happen with rate $\lambda_i = 1/T_i$ and to the transitional states with rate $\eta_i = 1/S_i$, where $i=\{R,V\}$.
For example, in the state where both populations are in the $+$ fixation state (the pink state $++$ in  Fig.~\ref{fig:telegraph}b,c), a beneficial mutation occurs in the virus with rate $\lambda_V$ moving it to a sweep state $+\downarrow$. 
The system moves to the discordant configuration $+-$ with effective rate $\eta_V$, until a beneficial mutation in the repertoire occurs with rate $\lambda_R$ and so on. 
In this setting, the turnover of a population in a time window $\Delta t$ is $1$ if the state at $t$ has a different fixation sign than the state at $t + \Delta t$, $0$ if the fixation sign is the same between the two time points (see section \ref{sec:SM_8state_MC} for more details).

This simplified Markov-chain qualitatively captures the correlations between turnovers of the full model in section~\ref{sec:pop_gen_model} (Fig.~\ref{fig:SM_8states}a and section \ref{sec:SM_8state_MC} of SM).
This case can be also solved analytically (Eq.~\ref{eq:SM_corr_toy}), allowing us to understand how the correlation regime depends on the population parameters and the timescales involved. 
In general, we see that we need selection to observe any correlation. Negative correlations, such as those observed in the data, require small or intermediate selection coefficients $s$, and small turnover time windows $\Delta t$ (Fig.~\ref{fig:telegraph}d and section \ref{sec:SM_corr_8state}).
The intuition behind the negative correlation is that a sweep in the second population is unlikely to occur before the sweep of the first population is almost complete, because the selective advantage for the second population is weaker during the sweep than after fixation. 
As a consequence, two sweeps are unlikely to be synchronous, implying an anti-correlation in the turnovers of the two populations when $\Delta t$ is of the order of the sweep time $S$.
Consistent with this reasoning, Fig.~\ref{fig:telegraph}d shows that the first maximum of negative correlation scales with the sweep time $S \sim \log (s N) / s$, which sets the timescale for negative correlations.

On the contrary, a longer $\Delta t$ will not be affected by this interference effect, and can lead to positive correlations when $\Delta t\sim T$, capturing the intuition that a sweep in a population favors a subsequent sweep in the other. 
Consistently, the maximum of positive correlation scales with the establishment time $T\sim 1/(s N \mu)$, determining the timescale of positive correlations (Fig.~\ref{fig:telegraph}d and  section \ref{sec:SM_corr_8state}). 
Increasing the time window $\Delta t$ after this scale, the first population has enough time to sweep back, generating a second region of negative correlation at larger $\Delta t$.
This alternating behavior of positive and negative correlations is damped as $\Delta t$ is furhter increased, eventually approaching zero correlation for $\Delta t \gg T$ (section \ref{sec:SM_corr_8state}).

Together, the simple model shows that the timescales of turnover determine the negative correlations observed in data. The results also suggest the limits of relatively weak selection coefficients and small turnover windows compared to mutation rates. 

\section{Discussion}
HIV and the immune system likely interact antagonistically in a co-evolutionary process, whose quantitative details are still not well understood.
Taking advantage of a dataset that tracks in time HIV sequences and B-cell repertoires, we show that this interaction exists and leaves traces at the population scale.
In particular, we find that HIV genetic turnover and lineage turnover are significantly negatively correlated.
Since lineage turnover is a measure that considers the immune system as a whole, this significant signal implies that the viral population interacts with a large number of distinct immune lineages or that the lineages involved have a big size.
A second surprising finding is that this correlation is negative, meaning that when the viral population slows down, the B-cell clones increase their rate of change (and vice-versa).

Using a simple model of co-evolution, we were able to show that negative correlations appear when the time delay for computing the turnover is comparable with the sweep time of mutations, while positive correlations emerge for delays of the order of the establishment time of the new mutant. The simplest one-locus model that we considered to derive these time scales clearly lacks biological realism, in particular in its neglecting mutational background and competition between lineages. Nevertheless, it shows that the observed negative correlations are driven by a specific interaction between the timescales of the problem and constrain the evolutionary regimes of the viral population and the repertoire. Specifically, it suggests that the viral population sweeps before the immune repertoire can respond.

While we cannot offer a detailed mechanism for how this evolutionary regime is obtained, we may speculate that the viral population within a sweep rapidly mutates away from the regions covered by the repertoire, so that the immune system cannot immediately adapt. 
Eventually the immune system catches up, but this happens on timescales where new beneficial mutations in the viral population may occur. 

By investigating the correlations between the two populations with a timeshift,
we found that a few measures of viral turnover and dN/dS negatively correlate with Ig measures one step forward in the future, while the opposite was never true.
This suggests that HIV evolution impacts the future state of the immune system.
One possible explanation for the absence of reciprocity could be that there is a difference in the time scales of response to changes in the other population.
Assume that the HIV virus responds fast to repertoire changes, while repertoires respond more slowly to changes in HIV composition.
Then the immune system would hold a longer memory of the past states of the viral population, leading to the observed delay in the anticorrelation.
Characterizing the time scales involved in these correlations could help us unveil crucial properties of the HIV immune-system interaction.
However, our attempts in this direction have been hampered by the limited size of the dataset.

The more complete model of HIV-Ig co-evolution with biologically reasonable parameters inspired by \cite{nourmohammad2016host} also reproduces the negative correlations in a wide range of biologically plausible parameters.
However, the model considers only the evolution of clonotypes within a single lineage and not the multi-lineage dynamics, since it is unclear what kind of dynamics and parameters to choose for modeling competition between lineages.
A future interesting line of research would be to better characterize this competitive dynamics, and inspect the correlations generated by the lineage turnover, which are the measures that show the strongest signal.

To conclude, in this work we have identified an unexpected pattern of anti-correlation in the longitudinal tracking of co-evolving HIV and Ig repertoires.
The approach we developped can help to analyze genetic data from other co-evolving populations with antagonist interactions, and to understand better the general rules that govern them.

\section*{Acknowledgements}
This work was partially supported by the
European Research Council Consolidator Grant n. 724208 and ANR-19-CE45-0018 ``RESP-REP" from the Agence Nationale de la Recherche.

\bibliographystyle{unsrt}

\begin{thebibliography}{10}

\bibitem{maclennan1994germinal}
Ian~CM MacLennan.
\newblock Germinal centers.
\newblock {\em Annual review of immunology}, 12(1):117--139, 1994.

\bibitem{allen2007germinal}
Christopher~DC Allen, Takaharu Okada, and Jason~G Cyster.
\newblock Germinal-center organization and cellular dynamics.
\newblock {\em Immunity}, 27(2):190--202, 2007.

\bibitem{campbell2013properties}
Catarina~D Campbell and Evan~E Eichler.
\newblock Properties and rates of germline mutations in humans.
\newblock {\em Trends in Genetics}, 29(10):575--584, 2013.

\bibitem{victora2012germinal}
Gabriel~D Victora and Michel~C Nussenzweig.
\newblock Germinal centers.
\newblock {\em Annual review of immunology}, 30:429--457, 2012.

\bibitem{shlomchik2012germinal}
Mark~J Shlomchik and Florian Weisel.
\newblock Germinal center selection and the development of memory b and plasma
  cells.
\newblock {\em Immunological reviews}, 247(1):52--63, 2012.

\bibitem{mesin2016germinal}
Luka Mesin, Jonatan Ersching, and Gabriel~D Victora.
\newblock Germinal center b cell dynamics.
\newblock {\em Immunity}, 45(3):471--482, 2016.

\bibitem{mcmichael2010immune}
Andrew~J McMichael, Persephone Borrow, Georgia~D Tomaras, Nilu Goonetilleke,
  and Barton~F Haynes.
\newblock The immune response during acute hiv-1 infection: clues for vaccine
  development.
\newblock {\em Nature Reviews Immunology}, 10(1):11--23, 2010.

\bibitem{fauci2003hiv}
Anthony~S Fauci.
\newblock Hiv and aids: 20 years of science.
\newblock {\em Nature medicine}, 9(7):839--843, 2003.

\bibitem{richman2003rapid}
Douglas~D Richman, Terri Wrin, Susan~J Little, and Christos~J Petropoulos.
\newblock Rapid evolution of the neutralizing antibody response to hiv type 1
  infection.
\newblock {\em Proceedings of the National Academy of Sciences},
  100(7):4144--4149, 2003.

\bibitem{moore2009limited}
Penny~L Moore, Nthabeleng Ranchobe, Bronwen~E Lambson, Elin~S Gray, Eleanor
  Cave, Melissa-Rose Abrahams, Gama Bandawe, Koleka Mlisana, Salim~S
  Abdool~Karim, Carolyn Williamson, et~al.
\newblock Limited neutralizing antibody specificities drive neutralization
  escape in early hiv-1 subtype c infection.
\newblock {\em PLoS pathogens}, 5(9):e1000598, 2009.

\bibitem{kwong2002hiv}
Peter~D Kwong, Michael~L Doyle, David~J Casper, Claudia Cicala, Stephanie~A
  Leavitt, Shahzad Majeed, Tavis~D Steenbeke, Miro Venturi, Irwin Chaiken,
  Michael Fung, et~al.
\newblock Hiv-1 evades antibody-mediated neutralization through conformational
  masking of receptor-binding sites.
\newblock {\em Nature}, 420(6916):678--682, 2002.

\bibitem{lyumkis2013cryo}
Dmitry Lyumkis, Jean-Philippe Julien, Natalia De~Val, Albert Cupo, Clinton~S
  Potter, Per-Johan Klasse, Dennis~R Burton, Rogier~W Sanders, John~P Moore,
  Bridget Carragher, et~al.
\newblock Cryo-em structure of a fully glycosylated soluble cleaved hiv-1
  envelope trimer.
\newblock {\em Science}, 342(6165):1484--1490, 2013.

\bibitem{simek2009human}
Melissa~D Simek, Wasima Rida, Frances~H Priddy, Pham Pung, Emily Carrow,
  Dagna~S Laufer, Jennifer~K Lehrman, Mark Boaz, Tony Tarragona-Fiol, George
  Miiro, et~al.
\newblock Human immunodeficiency virus type 1 elite neutralizers: individuals
  with broad and potent neutralizing activity identified by using a
  high-throughput neutralization assay together with an analytical selection
  algorithm.
\newblock {\em Journal of virology}, 83(14):7337--7348, 2009.

\bibitem{liao2013co}
Hua-Xin Liao, Rebecca Lynch, Tongqing Zhou, Feng Gao, S~Munir Alam, Scott~D
  Boyd, Andrew~Z Fire, Krishna~M Roskin, Chaim~A Schramm, Zhenhai Zhang, et~al.
\newblock Co-evolution of a broadly neutralizing hiv-1 antibody and founder
  virus.
\newblock {\em Nature}, 496(7446):469--476, 2013.

\bibitem{mccoy2017identification}
Laura~E McCoy and Dennis~R Burton.
\newblock Identification and specificity of broadly neutralizing antibodies
  against hiv.
\newblock {\em Immunological reviews}, 275(1):11--20, 2017.

\bibitem{walker2009broad}
Laura~M Walker, Sanjay~K Phogat, Po-Ying Chan-Hui, Denise Wagner, Pham Phung,
  Julie~L Goss, Terri Wrin, Melissa~D Simek, Steven Fling, Jennifer~L Mitcham,
  et~al.
\newblock Broad and potent neutralizing antibodies from an african donor reveal
  a new hiv-1 vaccine target.
\newblock {\em Science}, 326(5950):285--289, 2009.

\bibitem{walker2011broad}
Laura~M Walker, Michael Huber, Katie~J Doores, Emilia Falkowska, Robert
  Pejchal, Jean-Philippe Julien, Sheng-Kai Wang, Alejandra Ramos, Po-Ying
  Chan-Hui, Matthew Moyle, et~al.
\newblock Broad neutralization coverage of hiv by multiple highly potent
  antibodies.
\newblock {\em Nature}, 477(7365):466--470, 2011.

\bibitem{kwong2013broadly}
Peter~D Kwong, John~R Mascola, and Gary~J Nabel.
\newblock Broadly neutralizing antibodies and the search for an hiv-1 vaccine:
  the end of the beginning.
\newblock {\em Nature Reviews Immunology}, 13(9):693--701, 2013.

\bibitem{klein2013antibodies}
Florian Klein, Hugo Mouquet, Pia Dosenovic, Johannes~F Scheid, Louise Scharf,
  and Michel~C Nussenzweig.
\newblock Antibodies in hiv-1 vaccine development and therapy.
\newblock {\em Science}, 341(6151):1199--1204, 2013.

\bibitem{fischer2010transmission}
Will Fischer, Vitaly~V Ganusov, Elena~E Giorgi, Peter~T Hraber, Brandon~F
  Keele, Thomas Leitner, Cliff~S Han, Cheryl~D Gleasner, Lance Green, Chien-Chi
  Lo, et~al.
\newblock Transmission of single hiv-1 genomes and dynamics of early immune
  escape revealed by ultra-deep sequencing.
\newblock {\em PloS one}, 5(8):e12303, 2010.

\bibitem{henn2012whole}
Matthew~R Henn, Christian~L Boutwell, Patrick Charlebois, Niall~J Lennon,
  Karen~A Power, Alexander~R Macalalad, Aaron~M Berlin, Christine~M Malboeuf,
  Elizabeth~M Ryan, Sante Gnerre, et~al.
\newblock Whole genome deep sequencing of hiv-1 reveals the impact of early
  minor variants upon immune recognition during acute infection.
\newblock {\em PLoS pathogens}, 8(3):e1002529, 2012.

\bibitem{barton2016relative}
John~P Barton, Nilu Goonetilleke, Thomas~C Butler, Bruce~D Walker, Andrew~J
  McMichael, and Arup~K Chakraborty.
\newblock Relative rate and location of intra-host hiv evolution to evade
  cellular immunity are predictable.
\newblock {\em Nature communications}, 7(1):1--10, 2016.

\bibitem{zanini2015population}
Fabio Zanini, Johanna Brodin, Lina Thebo, Christa Lanz, G{\"o}ran Bratt, Jan
  Albert, and Richard~A Neher.
\newblock Population genomics of intrapatient hiv-1 evolution.
\newblock {\em Elife}, 4:e11282, 2015.

\bibitem{mouquet2014antibody}
Hugo Mouquet.
\newblock Antibody b cell responses in hiv-1 infection.
\newblock {\em Trends in immunology}, 35(11):549--561, 2014.

\bibitem{victora2018primary}
Gabriel~D Victora and Hugo Mouquet.
\newblock What are the primary limitations in b-cell affinity maturation, and
  how much affinity maturation can we drive with vaccination? lessons from the
  antibody response to hiv-1.
\newblock {\em Cold Spring Harbor perspectives in biology}, 10(5):a029389,
  2018.

\bibitem{kreer2022determining}
Christoph Kreer, Cosimo Lupo, Meryem~Seda Ercanoglu, Lutz Gieselmann, Natanael
  Spisak, Jan Grossbach, Maike Schlotz, Philipp Schommers, Henning Gruell,
  Leona Dold, et~al.
\newblock Determining probabilities of hiv-1 bnab development in healthy and
  chronically infected individuals.
\newblock {\em bioRxiv}, 2022.

\bibitem{hoehn2015dynamics}
Kenneth~B Hoehn, Astrid Gall, Rachael Bashford-Rogers, SJ~Fidler, S~Kaye,
  JN~Weber, MO~McClure, SPARTAC~Trial Investigators, Paul Kellam, and Oliver~G
  Pybus.
\newblock Dynamics of immunoglobulin sequence diversity in hiv-1 infected
  individuals.
\newblock {\em Philosophical Transactions of the Royal Society B: Biological
  Sciences}, 370(1676):20140241, 2015.

\bibitem{nourmohammad2019fierce}
Armita Nourmohammad, Jakub Otwinowski, Marta {\L}uksza, Thierry Mora, and
  Aleksandra~M Walczak.
\newblock Fierce selection and interference in b-cell repertoire response to
  chronic hiv-1.
\newblock {\em Molecular biology and evolution}, 36(10):2184--2194, 2019.

\bibitem{wang2015manipulating}
Shenshen Wang, Jordi Mata-Fink, Barry Kriegsman, Melissa Hanson, Darrell~J
  Irvine, Herman~N Eisen, Dennis~R Burton, K~Dane Wittrup, Mehran Kardar, and
  Arup~K Chakraborty.
\newblock Manipulating the selection forces during affinity maturation to
  generate cross-reactive hiv antibodies.
\newblock {\em Cell}, 160(4):785--797, 2015.

\bibitem{nourmohammad2016host}
Armita Nourmohammad, Jakub Otwinowski, and Joshua~B Plotkin.
\newblock Host-pathogen coevolution and the emergence of broadly neutralizing
  antibodies in chronic infections.
\newblock {\em PLoS genetics}, 12(7):e1006171, 2016.

\bibitem{molari2020quantitative}
Marco Molari, Klaus Eyer, Jean Baudry, Simona Cocco, and R{\'e}mi Monasson.
\newblock Quantitative modeling of the effect of antigen dosage on b-cell
  affinity distributions in maturating germinal centers.
\newblock {\em Elife}, 9:e55678, 2020.

\bibitem{strauli2019genetic}
Nicolas Strauli, Emily~Kathleen Fryer, Olivia Pham, Mohamed Abdel-Mohsen,
  Shelley~N Facente, Christopher Pilcher, Pleuni Pennings, Satish Pillai, and
  Ryan~D Hernandez.
\newblock The genetic interaction between hiv and the antibody repertoire.
\newblock {\em BioRxiv}, 2019.

\bibitem{hatada2010human}
Makiko Hatada, Kazuhisa Yoshimura, Shigeyoshi Harada, Yoko Kawanami, Junji
  Shibata, and Shuzo Matsushita.
\newblock Human immunodeficiency virus type 1 evasion of a neutralizing anti-v3
  antibody involves acquisition of a potential glycosylation site in v2.
\newblock {\em Journal of General Virology}, 91(5):1335--1345, 2010.

\bibitem{ringe2012unique}
Rajesh Ringe, Lipsa Das, Ipsita Choudhary, Deepak Sharma, Ramesh Paranjape,
  Virander~Singh Chauhan, and Jayanta Bhattacharya.
\newblock Unique c2v3 sequence in hiv-1 envelope obtained from broadly
  neutralizing plasma of a slow progressing patient conferred enhanced virus
  neutralization.
\newblock {\em PLoS One}, 2012.

\bibitem{nei1986simple}
Masatoshi Nei and Takashi Gojobori.
\newblock Simple methods for estimating the numbers of synonymous and
  nonsynonymous nucleotide substitutions.
\newblock {\em Molecular biology and evolution}, 3(5):418--426, 1986.

\bibitem{yang2000statistical}
Ziheng Yang and Joseph~P Bielawski.
\newblock Statistical methods for detecting molecular adaptation.
\newblock {\em Trends in ecology \& evolution}, 15(12):496--503, 2000.

\bibitem{tas2016visualizing}
Jeroen~MJ Tas, Luka Mesin, Giulia Pasqual, Sasha Targ, Johanne~T Jacobsen,
  Yasuko~M Mano, Casie~S Chen, Jean-Claude Weill, Claude-Agn{\`e}s Reynaud,
  Edward~P Browne, et~al.
\newblock Visualizing antibody affinity maturation in germinal centers.
\newblock {\em Science}, 351(6277):1048--1054, 2016.

\bibitem{lemey2006hiv}
Philippe Lemey, Andrew Rambaut, and Oliver~G Pybus.
\newblock Hiv evolutionary dynamics within and among hosts.
\newblock {\em Aids Rev}, 8(3):125--140, 2006.

\bibitem{elhanati2015inferring}
Yuval Elhanati, Zachary Sethna, Quentin Marcou, Curtis~G Callan~Jr, Thierry
  Mora, and Aleksandra~M Walczak.
\newblock Inferring processes underlying b-cell repertoire diversity.
\newblock {\em Philosophical Transactions of the Royal Society B: Biological
  Sciences}, 370(1676):20140243, 2015.

\bibitem{neher2013genetic}
Richard~A Neher.
\newblock Genetic draft, selective interference, and population genetics of
  rapid adaptation.
\newblock {\em arXiv preprint arXiv:1302.1148}, 2013.

\bibitem{desai2007beneficial}
Michael~M Desai and Daniel~S Fisher.
\newblock Beneficial mutation--selection balance and the effect of linkage on
  positive selection.
\newblock {\em Genetics}, 176(3):1759--1798, 2007.

\bibitem{ebert2020host}
Dieter Ebert and Peter~D Fields.
\newblock Host--parasite co-evolution and its genomic signature.
\newblock {\em Nature Reviews Genetics}, 21(12):754--768, 2020.

\bibitem{sievers2011fast}
Fabian Sievers, Andreas Wilm, David Dineen, Toby~J Gibson, Kevin Karplus,
  Weizhong Li, Rodrigo Lopez, Hamish McWilliam, Michael Remmert, Johannes
  S{\"o}ding, et~al.
\newblock Fast, scalable generation of high-quality protein multiple sequence
  alignments using clustal omega.
\newblock {\em Molecular systems biology}, 7(1):539, 2011.

\bibitem{vander2014presto}
Jason~A Vander~Heiden, Gur Yaari, Mohamed Uduman, Joel~NH Stern, Kevin~C
  O’Connor, David~A Hafler, Francois Vigneault, and Steven~H Kleinstein.
\newblock presto: a toolkit for processing high-throughput sequencing raw reads
  of lymphocyte receptor repertoires.
\newblock {\em Bioinformatics}, 30(13):1930--1932, 2014.

\bibitem{gupta2015change}
Namita~T Gupta, Jason~A Vander~Heiden, Mohamed Uduman, Daniel Gadala-Maria, Gur
  Yaari, and Steven~H Kleinstein.
\newblock Change-o: a toolkit for analyzing large-scale b cell immunoglobulin
  repertoire sequencing data.
\newblock {\em Bioinformatics}, 31(20):3356--3358, 2015.

\bibitem{gupta2017hierarchical}
Namita~T Gupta, Kristofor~D Adams, Adrian~W Briggs, Sonia~C Timberlake,
  Francois Vigneault, and Steven~H Kleinstein.
\newblock Hierarchical clustering can identify b cell clones with high
  confidence in ig repertoire sequencing data.
\newblock {\em The Journal of Immunology}, 198(6):2489--2499, 2017.

\end{thebibliography}

\pagebreak

\onecolumngrid

\newpage

\begin{center}
\textbf{\large SUPPLEMENTARY MATERIAL}
\end{center}

\setcounter{figure}{0}
\setcounter{section}{0}
\setcounter{table}{0}
\setcounter{equation}{0}
\renewcommand{\figurename}{Supplementary Figure}
\renewcommand{\thefigure}{S\arabic{figure}}
\renewcommand{\thesection}{S\arabic{section}}
\renewcommand{\tablename}{Supplementary Table}
\renewcommand{\thetable}{S\arabic{table}}
\renewcommand{\theequation}{S\arabic{equation}}

\section{Dataset description}
\label{sec:SM_dataset_descr}

The open access database of Strauli et al~\cite{strauli2019genetic} is available om the Sequence Read Archive, SRA, under the BioProject ID PRJNA543982.
\url{https://www.ncbi.nlm.nih.gov/bioproject/?term=PRJNA543982}.
To download and process the dataset, the repository \url{https://github.com/statbiophys/HIV_coevo.git} contains all the necessary scripts.

\begin{figure}[h]
    \centering
    \includegraphics[width=\textwidth]{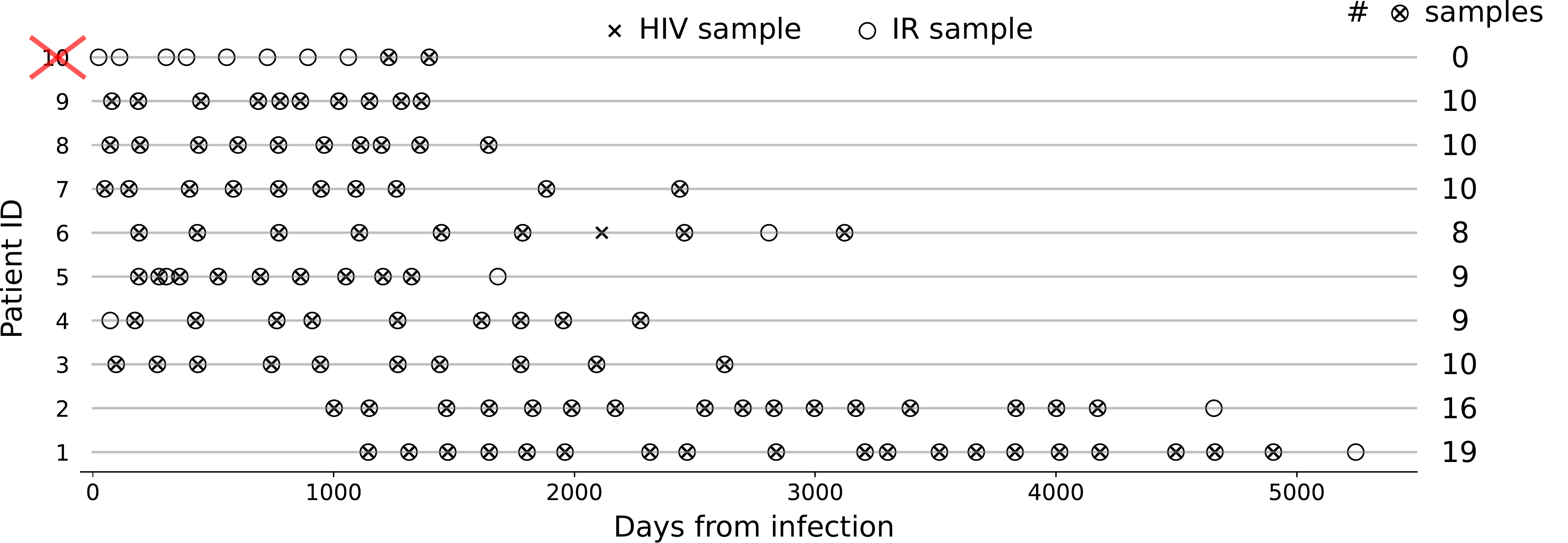}
    \caption{Distribution of HIV and immune repertorie samples along the estimated time from infection.
    The last column of numbers shows how many time points with both HIV and immune repertoire samples there are for each patient.
    Patient 10 has been excluded from the analysis.}
    \label{fig:SM_data_descr}
\end{figure}

As already mentioned in the main text, the dataset is composed of longitudinal samples of 10 patients.
Fig.~\ref{fig:SM_data_descr} shows the temporal configuration of the HIV and immune repertoire samples.
The samples are considered in the analysis if both the HIV and the immune repertoire are present at the same time point, allowing us to compare the evolutionary dynamics of the two systems.
Since patient 10 has only two available time points, and therefore the turnover trajectory is composed of only one point, it was excluded from the analysis.
The last time point of patient 1, 2 and 5 are taken after administration of antiretroviral therapy. Those samples are excluded from our analysis as they do not have any HIV sequence.

\section{Dataset processing}
\label{sec:SM_dataset_process}

\subsection{Assembling the HIV sequences}

The HIV sequence pairs are assembled from the fastq files using a python notebook script contained in the linked repository.
For each read and its reverse complement, the overlapping region between the two is found through a local alignment with a large gap penalty.
The two reads are then assembled together. 
If at one position of the overlap the two sequences have different bases, the one with highest quality score is kept.

Some quality checks are then performed during this procedure.
Assembled sequences with an average quality score smaller than 32 are discarded. 
 
Sequences with length smaller than 400bp and larger than 425bp are also discarded (which is the typical range of the \textit{env} gene length).

The list of identical sequences obtained in this way are then counted, and the final output of this procedure is the list of unique sequences with their count.
We discarded sequences with a single read (singletons).
We use only sequences that occur at least twice and, therefore, have a very small probability of having sequencing errors.

At the end of this procedure, two samples, patient 6 at day 2808 and patient 2 at day 4656, were discarded because almost none of the reads passed the quality checks (Fig.~\ref{fig:SM_data_descr} does not show these two samples).

\subsection{Framing  and multiple aligning of the HIV sequences}

In order to track Single Nucleotide Polymorphism (SNP) along the different time samples of a patient, the sequences at any time point need to be aligned together.
To distinguish between synonymous and non-synonymous mutations the sequences need to be framed, such that it is possible to identify the corresponding codons in the sequence of nucleotides.
These two operations are performed in a python notebook contained in the repository.

The framing operation is performed first.
For each sequence, the number of stop codons is counted for the three possible choices of the frame.
The frame that minimises the number of stop codons is then chosen.
The number of stop codons in such a frame is most of the time zero, as expected from productive sequences.
However, a small fraction of sequences with few stop codons remains.
The ones with more than one count of stop codons are discarded.

The multiple alignment of sequences among different time samples of a given patient is performed with Clustal Omega \cite{sievers2011fast}.
To preserve the codon structure in the multiple alignment, this operation is performed in the amino-acid space.
Therefore, each sequence is converted from nucleotides to amino-acids, then all the sequences of all the time points are aligned, and finally the gap-structure of the alignment is mapped back to the nucleotide sequences.

\subsection{Assembling and blasting the IgH clonotypes}

The creation of clonotypes from the immunoglobulin heavy chains reads has been done with PRESTO \cite{vander2014presto} and IgBlast \cite{gupta2015change}.
The sequence of operations performed in the script are:
\begin{itemize}
    \item A quality filter, which discards reads with average quality score less than 30.
    \item Assembling each read with its reverse complement.
    \item Collapsing the same reads, which also defined a counter of multiplicity.
    \item Blasting the sequence against the database of human antibodies and identifying the proper V gene, J gene, and CDR3.
\end{itemize}

A list of clonotypes identified by a given V-gene, J-gene and CDR3 sequence is then obtained for each sample.

\subsection{Building the IgH lineages}
\label{sec:SM_lineages}

\begin{figure}
\centering
\includegraphics[width=0.5\textwidth]{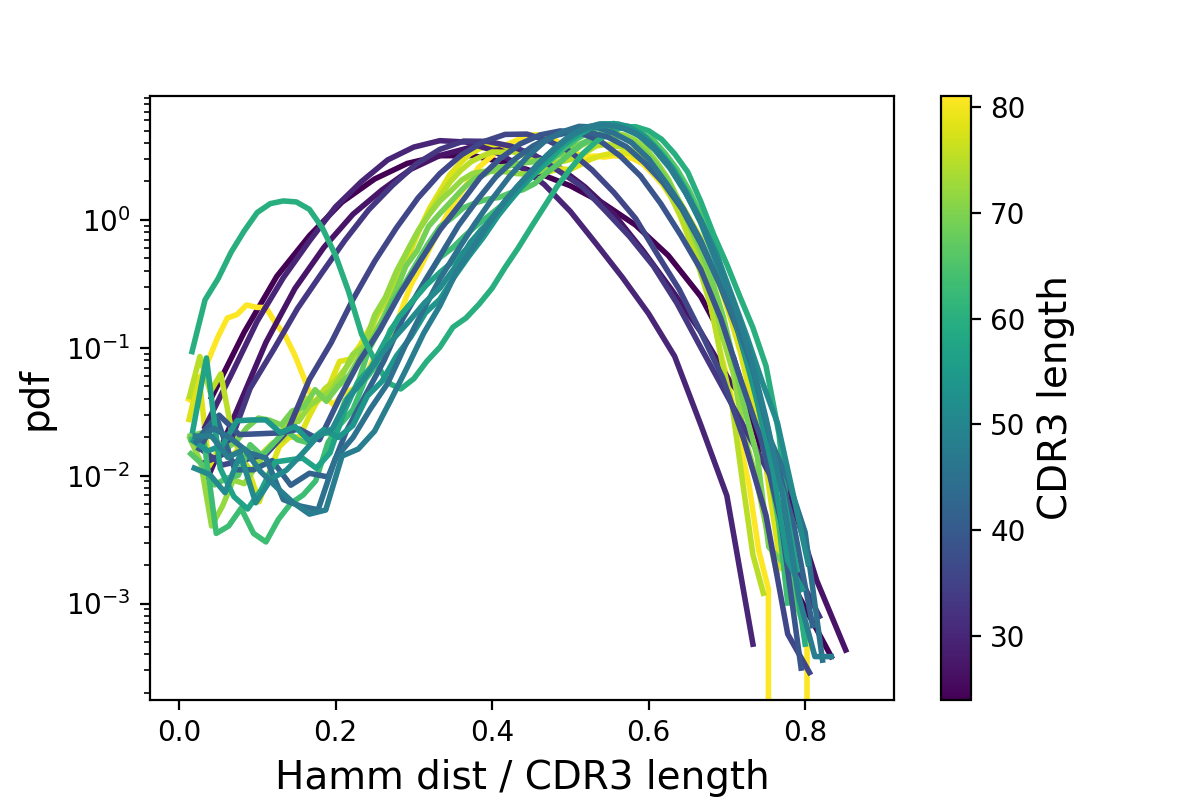}
\caption{Hamming distance divided by CDR3 length distribution among the pairs of CDR3 sequences at given a length.
The color represents the length of the CDR3.}
\label{fig:hamm_dist_given_length}
\end{figure}

For each patient, the IgH clonotypes were clustered into lineages/families.
The procedure was performed over all the clonotypes at any time point for each patient.
They were first grouped together in different classes defined by a specific V gene, J gene (annotated with IgBlast at the previous step) and CDR3 length.
We call these groups VJL classes.
We discarded all the classes having length smaller than 30bp and also all the clonotypes having count equal to 1 (singletons).

The distribution of Hamming distance of pairs of CDR3 sequences (divided by the CDR3 length) is shown in Fig.~\ref{fig:hamm_dist_given_length}.
It is composed of a bulk of pairs at large distances (which is compatible with a null recombination model) and a second small peak, typically at zero.
The distances in the small peak are likely due to the very similar pairs created during the somatic expansion in the germline centers, and therefore belonging to the same lineages.
This plot gives us a rough idea about the typical Hamming distance that needs to be chosen to cluster clonotypes in the same family (approximately at the minimum between the two modes), which we choose to be 0.1.
We perform single linkage clustering within each VJL class with that threshold.
This procedure and threshold are similar to the one suggested in \cite{gupta2017hierarchical} and used in \cite{nourmohammad2019fierce}.
The clustering is performed with a very fast in-house prefix tree algorithm (\url{https://github.com/statbiophys/ATrieGC})
The resulting clusters define the B-cell repertoire families/lineages.

\subsection{Computing Single Nucleotide Polymorphisms}
\label{sec:SM_SNPs}

Given a list of multiple-aligned sequences, the Single Nucleotide Polymorphisms (SNP) are computed with respect to a reference sequence,
which is chosen as the consensus sequence among the sequences at the first time point where the lineage appeared.

Each SNP is identified its position and the new nucleotide. 
Since the multiple alignment generates gaps, gaps are treated in the same manner as nucleotides and  SNPs of gaps are also computed.
We keep track of two quantities of a given SNP.
The first is the frequency, defined as the fraction of unique sequences in which it appears among all the unique sequences at a given time point.
Note that the sequence counts are not taken into account.
The second quantity is the abundance, i.e. the sum of the sequence counts of all the sequences in which the SNP appears.

Synonymous mutations are SNP that change the amino acid sequence.
Gap SNPs are always considered non-synonymous.
The normalization coefficients $k$ of Eq.~\ref{eq:dn/ds} of the main text is computed in the following way.
For each codon $c$ of the reference sequence, all the possible $9$ mutations are considered and the number of synonymous, $s_c$, and non-synonymous changes, $n_c$, are counted.
The coefficient $k$ is then:
\begin{equation}
k = \frac{\sum_c s_c}{\sum_c n_c} .
\end{equation}

\begin{figure}
\centering
\includegraphics[width=0.5\textwidth]{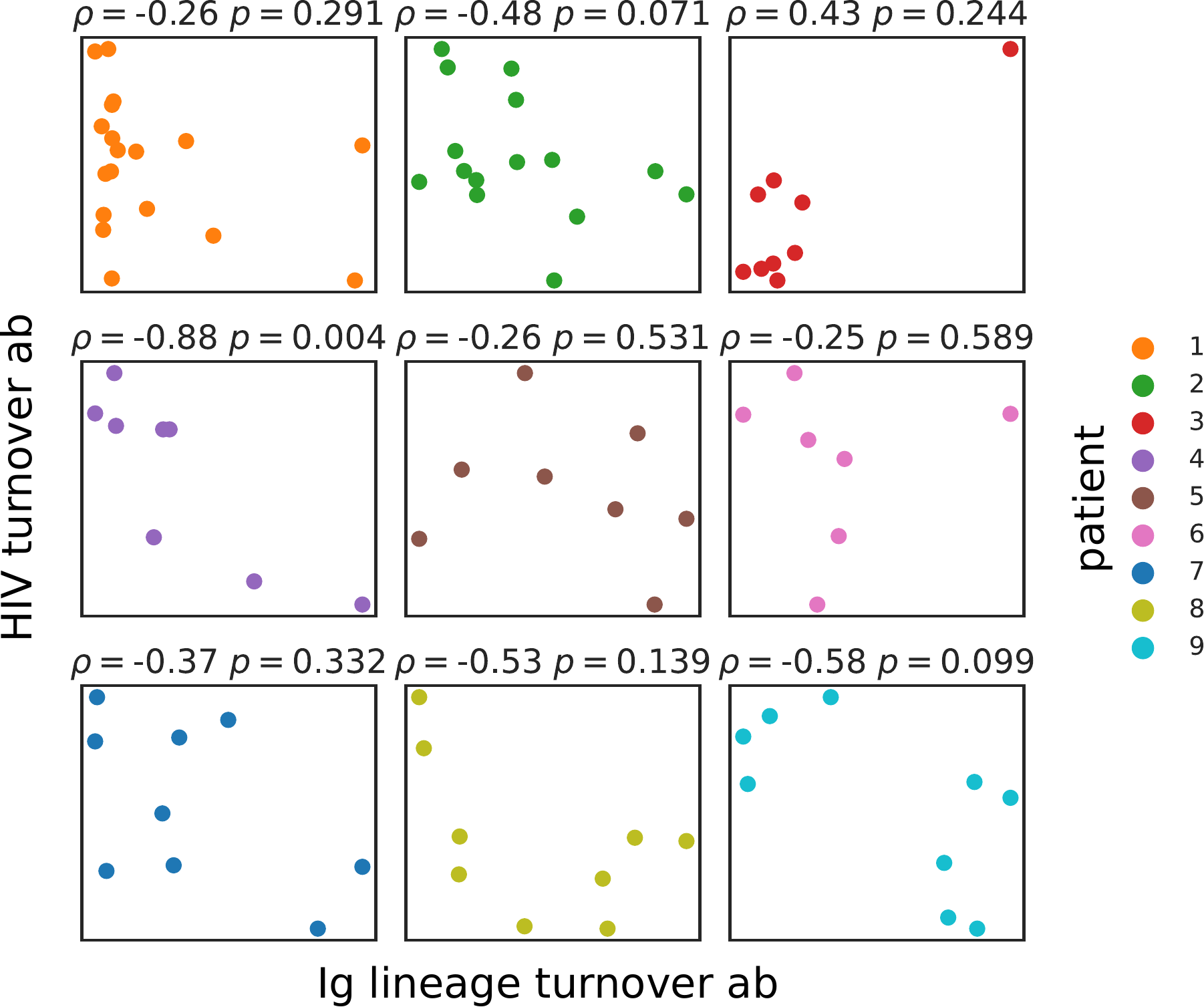}
\caption{Scatter plots of the trajectories of \textit{Ig lineage turnover ab} and \textit{HIV turnover ab} for the 9 patients of the dataset.
The Spearman correlation coefficient $\rho$ and the correlation p-value $p$ is reported as the title of each plot.}
\label{fig:SM_corr_single_pat}
\end{figure}

\begin{figure}
\centering
\includegraphics[width=0.8\textwidth]{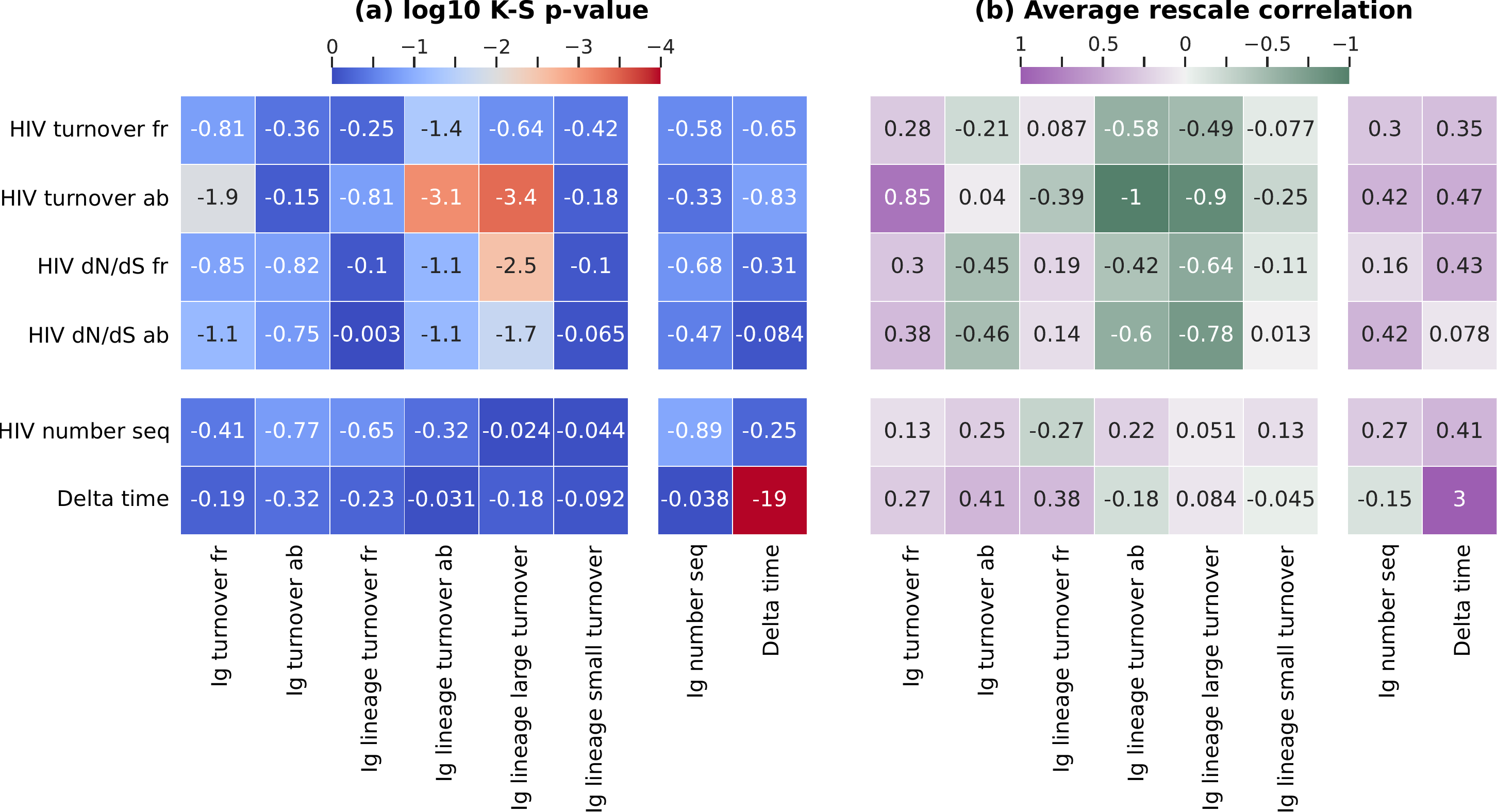}
\caption{(a) The base 10 logarithm of the p-value of the Kolmogorov -Smirnov test shown in Fig.~\ref{fig:ks_stat_test} of the main text. This number is computed for each pair of HIV and Ig measures introduced in the main text.
In addition, the correlation with the number of sequences (in the two time point of the variation) of HIV and Ig and the length of the time interval between the two time points is tested against every measure.
(b) The average rescaled correlation coefficient of each pair of trajectories. }
\label{fig:SM_ks_p_all}
\end{figure}

\begin{figure}
\centering
\includegraphics[width=0.7\textwidth]{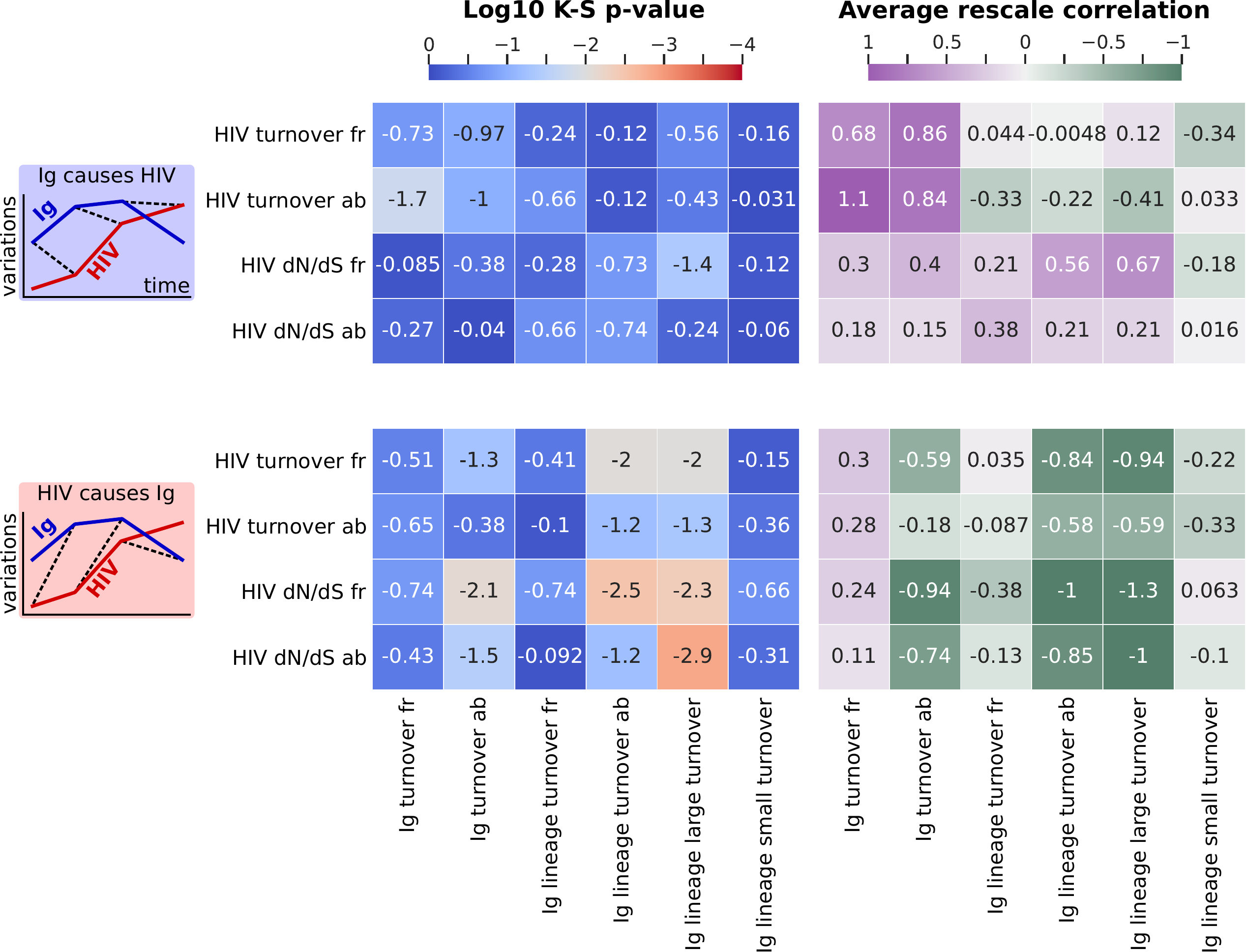}
\caption{Same plots for Fig.~\ref{fig:SM_ks_p_all} for the temporal shifts of the trajectories.}
\label{fig:SM_ks_p_all_shift}
\end{figure}

\section{Null model with internal correlations}
\label{sec:SM_null_corr}

\begin{figure}
\centering
\includegraphics[width=0.9\textwidth]{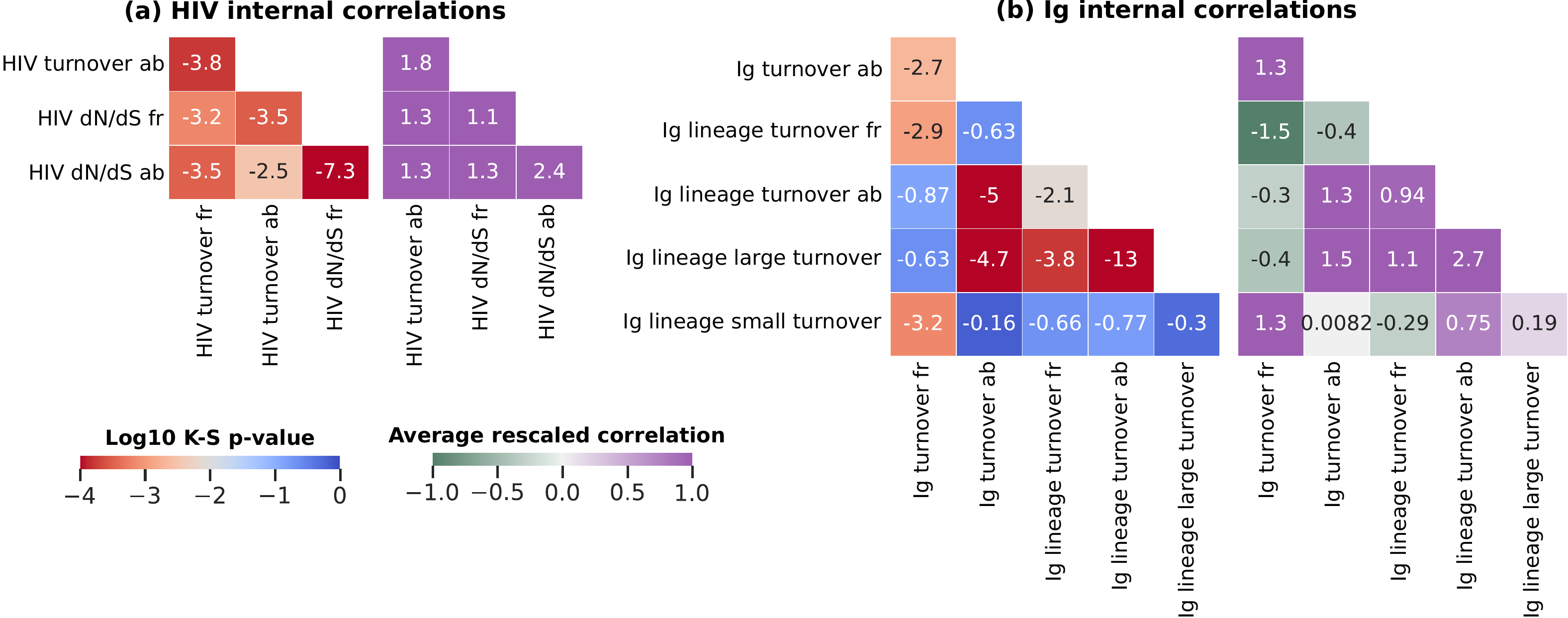}
\caption{Same plots of Fig.~\ref{fig:SM_ks_p_all} for the measures within the same population.}
\label{fig:SM_ks_p_internal}
\end{figure}

In this section we address the problem of assessing the global significance of of the list of p-values obtained from the Kolmogorov Smirnov tests of Fig.~\ref{fig:SM_ks_p_all} and \ref{fig:SM_ks_p_all_shift}.
Two facts must be taken into account: the multiple testing of the 24 pairs and the internal correlations between variables of HIV and between the ones of the immune system, see Fig.~\ref{fig:SM_ks_p_internal}.
In order to do so, we generate an ensemble of null scenarios, where the outcome of each one is, for each patient, a set of 4 trajectories for the HIV measures (table \ref{tab:hiv_measures}) and 6 trajectories for the immune system (table \ref{tab:abr_measures}).
The 4 HIV trajectories contain the same internal correlation (on average) as the real ones. The same is true for the 6 trajectories of the immune system.
However, each pair of HIV-Ig trajectories is uncorrelated. 
To this end, for each patients having $n$ time points, $n-1$\footnote{We need to generate trajectories of the variation measures, that are composed of $n-1$ time points for a patient sampled $n$ times.} four-dimensional vectors are extracted  from a multivariate random distribution, whose covariance matrix is fixed by the empirical internal correlation of the HIV measures. 
The same procedure is applied to the immune repertoires, where the multivariate probability distribution is now six-dimensional.
We tried to build this model using a multivariate Gaussian or a multivariate log-normal. Both distributions lead to the same result because the Spearman correlation is insensitive to how the points are distributed.

\begin{figure}
\centering
\includegraphics[width=0.4\textwidth]{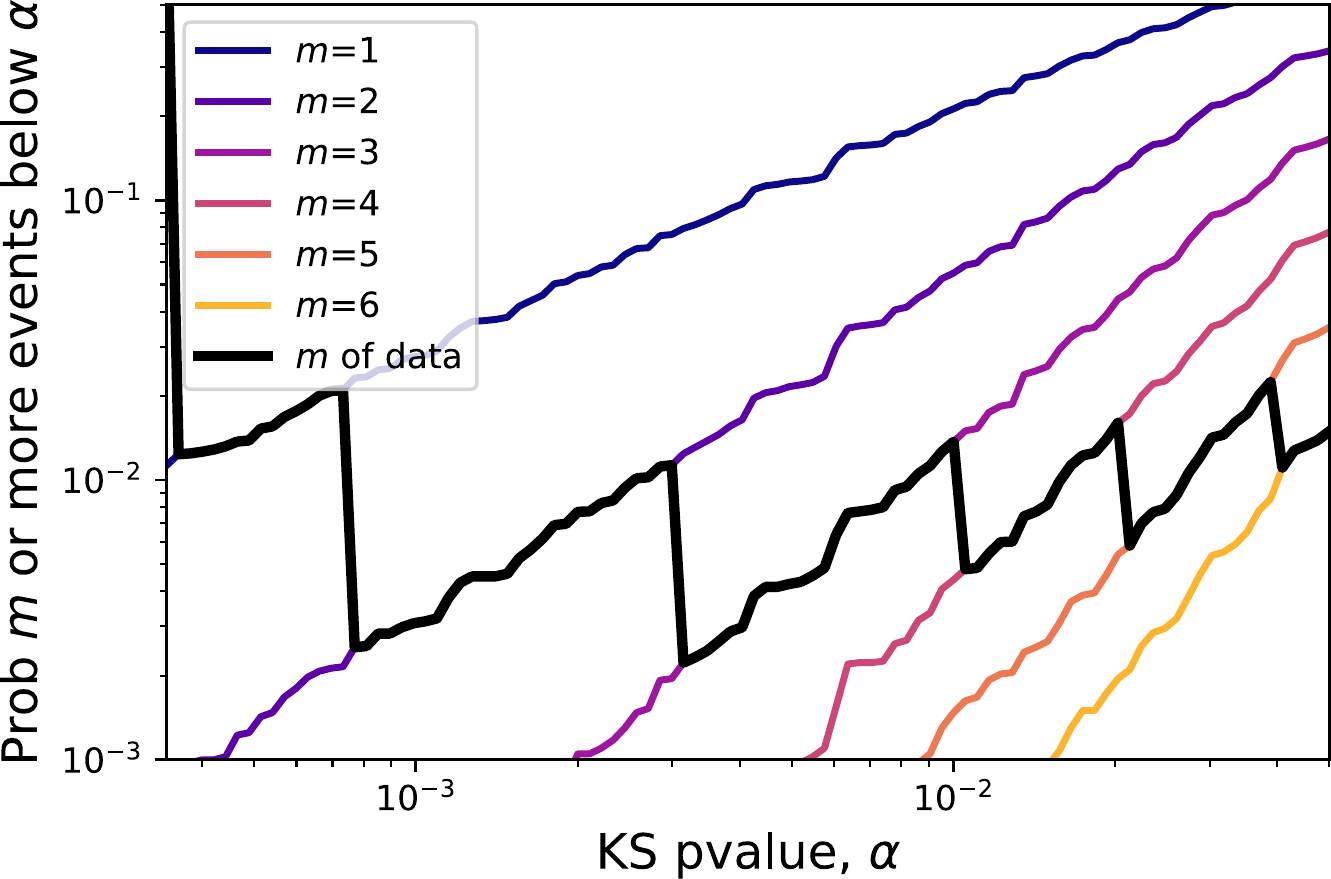}
\caption{Same plots as Fig.~\ref{fig:SM_ks_p_all} for the measures within the same population.}
\label{fig:SM_nullcorr}
\end{figure}

A large number of these scenarios is generated, and for each one of them, the 24 Kolmogorov Smirnov p-values are computed.
One can then calculate the fraction of scenarios having $m$ or more significant pairs with a K-S p-value smaller than $\alpha$.
In Fig.~\ref{fig:SM_nullcorr}, this probability is shown as a function of $\alpha$ and for different $m$, i.e. lines with different colors.
On top of those lines, the black line show the probability that, at given a $\alpha$, the null scenarios generate the same number or more of significant pairs as in the data.
For example, at p-value $\alpha=0.001$, in the data there are 2 significant pairs, and the probability that the null scenario generates this or a more extreme configuration is around $0.002$.
The message of Fig.~\ref{fig:SM_nullcorr} is  that it is very unlikely to generate null scenarios that preserve internal correlations with the same number of significant pairs as in the data.
In particular, going from a significance $\alpha$ on the KS p-value from $0.05$ to around $0.0001$ the probability of generating a null scenario as significant as the data (or more significant) is always around or smaller than $0.01$.

\section{Wright Fisher simulation of the co-evolving binary strings}
\label{sec:SM_WF}

The co-evolutionary dynamics is simulated using the Wright-Fisher model: a population genetics model with non-overlapping generations.
The code has been written in C++ and it is available in the repository  \url{https://github.com/statbiophys/HIV_coevo.git}.
The algorithmic steps are the following:
\begin{itemize}
	\item The two populations with size $N_V$ and $N_R$, are initialized with $5$ random sequences of binary variables of length $L$, each one with abundance $N_V/5$, or $N_R/5$.
	\item  For each genotype, characterized by a given sequence of binary variables, we compute its affinity to all the sequences of the other population (Eq.~\ref{eq:affinity}). The affinities are stored in the memory to avoid the affinity computations at each replication round.
	This memory is updated every time new sequences enter the process.
	\item The following process starts and its iterated for a given number of generations $T$. At each generation:
	\begin{itemize}
		\item Given the stored affinities, the fitness for each gentoype, $f_i$, is computed, Eq.~\ref{eq:linear_fitness}.
		\item We perform a multinomial sampling for each population. We generate $N_V$ or $N_R$ sequences, where each genetype has a probability proportional to $\exp(f_i) x_i$ of being sampled ($x_i = n_i/N$ is the frequency of the genotype).
		\item Each obtained genotype generates a number of  mutations that follows a binomial distribution with probability $\mu_R L$ (or $\mu_V L$) and $n_i$ samples, where $\mu_{R/V}$ is the mutation rate per site, per sequence, per generation.
		In this way $\mu N L$ mutations are generated on average in each population.
		\item Each mutation causes the sign change of one variable of the corresponding genotype (chosen at random among the $L$ variables).
		\item The genotype affinities are computed against the new mutants and stored in memory.
	\end{itemize}
\end{itemize}

In general, the process is repeated for $5,000$ generations to converge to steady state and then the observables are computed from that point on.

In the case of $L=50$, the simulations assume that each new mutation generates always a new genotype.
In the very unlikely case of generating a mutant that it is equal to an already existing one, it is treated as a different genotype with its own abundance.
We made this choice because it is computationally expensive to check at every round of mutations if the new mutants are really new, and since it is very unlikely that this happens, we expect that such events do not have relevant consequences for our results.
We treat differently the case of small $L$ (like the case discussed in section \ref{sec:telegraph} for $L=1$), where instead we check if every new mutant is already present.

\section{Discrete Markov-Chain for the sweep-regime of co-evolving populations of binary variables}
\label{sec:SM_8state_MC}

\subsection{Definition of the model and computation of the propagators}

The discrete Markov Chain introduced in the main text is composed of $8$ states $\sigma = \sigma_R \sigma_V\in \lbrace ++, +\downarrow, +-, \downarrow-, --, -\uparrow, -+, \uparrow+ \rbrace$.
The notation indicates the fixation sign for the B-cell receptors, $\sigma_R$, and the one for viral population, $\sigma_V$ (see Fig.~ \ref{fig:telegraph}).
For example, $\sigma_R \sigma_V = +-$ is a configuration in which the receptors all have variable $+1$ while the virus $-1$.
The tilde put on the top of the sign defines a switching state.
For example, $\sigma_R \sigma_V = \downarrow-$ is the state having the immune receptors that are progressively aligning with the viruses in the down state.
Note that each state can only move to only one single other state, and the rates at which this happens is contained in the following transition matrix:
\begin{equation}
W =\begin{pmatrix}
-\lambda_V & 0 & 0 & 0 & 0 & 0 & 0 & \eta_R \\
\lambda_V & -\eta_V & 0 & 0 & 0 & 0 & 0 & 0 \\
0 & \eta_V & -\lambda_R & 0 & 0 & 0 & 0 & 0 \\
0 & 0 & \lambda_R & -\eta_R & 0 & 0 & 0 & 0 \\
0 & 0 & 0 & \eta_R & -\lambda_V & 0 & 0 & 0 \\
0 & 0 & 0 & 0 & \lambda_V & -\eta_V & 0 & 0 \\
0 & 0 & 0 & 0 & 0 & \eta_V & -\lambda_R & 0 \\
0 & 0 & 0 & 0 & 0 & 0 & \lambda_R & -\eta_R 
\end{pmatrix}
\end{equation}
where $\lambda = 1/T \sim s N \mu$ is the inverse of the typical establishment time of a mutation, and $\eta = 1/S \sim s/\log(N s)$ the inverse of the switch time.

The temporal dynamics of the Markov Chain are described by the master equation for the probability vector $\bar{P}(t) = (p_1(t), \ldots, p_8(t))^T$, which reads $\partial_t \bar{P}(t) = W \bar{P}(t)$.
The equation above can be solved, and the solution has the form:
\begin{equation*}
\bar{P}(t) = \exp(W t) \bar{P}(0) = \sum_{n=0}^\infty \frac{W^n t^n}{n!}\bar{P}(0)  = U \sum_{n=0}^\infty \frac{D^n t^n}{n!} U^{-1} \bar{P}(0) = U \exp(D t) U^{-1} \bar{P}(0)
\end{equation*}
where $U$ is the matrix having the eigenvectors of $W$ as columns, $D$ the diagonal matrix having the eigenvalues $\lambda_i$ as elements, and $(\exp(D t))_{ij} = \exp(\lambda_i t)\delta_{ij}$.
From this expression we calculate the propagator from state $\sigma$ at time $0$ to $\sigma'$ at time $t$, $p(\sigma',t|\sigma,0)$.

\subsection{Computing the turnover correlation}

We are interested in computing the correlation between turnover of the two populations.
In this simple setting, we make an approximation and say that the turnover between a time $t$ and a time $t + \Delta t$ can be only 0 or 1.
Specifically, it is 1 if $\sigma_R$ (or $\sigma_V$) changes in that interval of time.
We consider the switching states having the same constant sign of the initial fixed state.
For example, if at time $t$ the state is $++$ and at time $t + \Delta t$ is $\downarrow-$, the first population has turnover 0 (we consider $+$ and $\downarrow$ as the same), while the second one has turnover 1.
This crude simplification is necessary for making analytical calculations.

It is convenient to introduce the stochastic variables $X_R(t,\Delta t)$ and $X_V(t,\Delta t)$ which are $1$ if the population has switched between $t$ and $t+\Delta t$ of $0$ otherwise.
In the following we will consider the system at stationary state, so the dependency on $t$ will be omitted.
The turnover correlation that we want to compute is then the correlation between these two variables, $C(\Delta t) = (\langle X_R(\Delta t)X_V(\Delta t) \rangle - \langle X_R(\Delta t) \rangle\langle X_V(\Delta t) \rangle) / \sigma(X_R(\Delta t)) / \sigma(X_V(\Delta t))$.
We can call $P_{R}(\Delta t) = \text{Prob}(X_{R}(\Delta t) = 1)$ the probability of having a turnover of $1$ in this time window in the receptor population (and similarly for $P_{V}(\Delta t)$), and $P_{RV}(\Delta t)$ the probability that both the virus and the receptors have turnover $1$.
The correlation function of these binary variables can then be written as follows:
\begin{equation*}
C(\Delta t) = \frac{P_{RV}(\Delta t) - P_R(\Delta t)P_V(\Delta t)}{\sqrt{P_R(\Delta t)(1-P_R(\Delta t))P_V(\Delta t)(1-P_V(\Delta t))}}.
\end{equation*}

The next step is to find the expression for the probabilities above.
By knowing the stationary probability of each state, $P_{\text{stat}}(\sigma)$ and the propagators $p(\sigma',t|\sigma,0)$, the turnover probability of the receptors reads:
\begin{equation*}
P_R(\Delta t) = \sum_{\sigma_R \sigma_V} P_{\text{stat}}(\sigma_R \sigma_V) \sum_{\sigma_V'} \left[ p(-\sigma_R \sigma_V',\Delta t|\sigma_R \sigma_V,0) + p(-\tilde{\sigma}_R \sigma_V',\Delta t|\sigma_R \sigma_V,0) \right],
\end{equation*}
where, in the case of $\sigma_R = \downarrow$ (or $\sigma_R = \uparrow$) we consider $\tilde{\sigma}_R = +$ (or $\tilde{\sigma}_R = -$).
Note that some combinations of $\sigma_R \sigma_V$ are not allowed, e.g. $\uparrow-$. Therefore the first summation is over the $8$ allowed states and the propagators with not-allowed arriving states are set to zero.
The final explicit expression leads to 32 terms to compute.
Similarly one can derive the expression for $P_V(\Delta t)$.
The last object we have to compute is
\begin{equation*}
\begin{split}
P_{RV}(\Delta t) = \sum_{\sigma_R \sigma_V} P_{\text{stat}}(\sigma_R \sigma_V) \left[ p(-\sigma_R -\sigma_V,\Delta t|\sigma_R \sigma_V,0) + p(-\tilde{\sigma_R} -\sigma_V,\Delta t|\sigma_R \sigma_V,0) \right.\\
\left. + p(-\sigma_R -\tilde{\sigma_V},\Delta t|\sigma_R \sigma_V,0) + p(-\tilde{\sigma_R} -\tilde{\sigma_V},\Delta t|\sigma_R \sigma_V,0) \right],
\end{split}
\end{equation*}
where, again, some propagators would lead to impossible states and are set to zero.
This quantity will contain 16 terms.

\begin{figure}
\centering
\includegraphics[width=\textwidth]{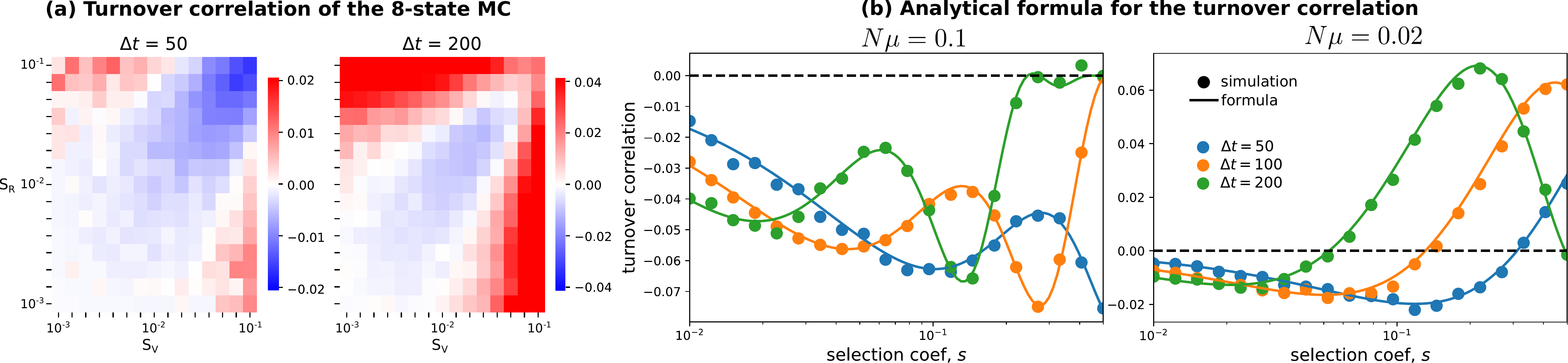}
\caption{Turnover correlations in the toy model for the successional mutations regime.
The first panel is a numerical simulation for different values of selection, $N = 10^3$ and $\mu = 10^3/50$.
The two plots are for different turnover time windows.
Panel (b) shows the comparison between simulations and Eq.~\ref{eq:SM_corr_toy}.}
\label{fig:SM_8states}
\end{figure}

Fig.~\ref{fig:SM_8states}a shows the turnover computed in this way over a large ensembles of simulated Markov Chains.
Using the same parameters of the full model, section \ref{sec:pop_gen_model}, the heatmap that emerges captures some features of the full model, Fig.~\ref{fig:bin_str_corr}, in particular for $s_R=s_V=s$.
Indeed, the full model generates stronger anti-correlations for different selection coefficients that are not captured by the Markov Chain. 
But, for equal selection, i.e. considering the diagonal of the map, the correlation moves from zero to negative correlation as $s$ increases to positive correlations for large $s$.
This pattern is present in both the models, and also the dependency on the turnover time window $\Delta t$ is similar: a larger $\Delta t$ moves this pattern towards smaller selection coefficients.

\subsection{Solution of the case with equal populations}
\label{sec:SM_corr_8state}

The general case with four different rates for the waiting and the establishment times of the two populations leads to a Markov process whose eigenvalues cannot be explicitly computed (they are implicit solutions of fourth-degree polynomials).
Therefore, we show the explicit solution for two populations with the same rates: $\lambda = 1/T_R = 1/T_V = s N \mu$, and for the waiting period, $\eta = 1/S_R = 1/S_V = s/\log(sN)$.

Explicit eigenvectors and eigenvalues are found, and explicit solution of the propagator are obtained, $p(i,t|j,0) = \bar{P}(t)_{ij}$.

By using the stationary distribution and the propagators, we compute the turnover correlation as shown in the previous section.
This long computation leads to the following expression:
\begin{equation}
C(t) = \frac{e^{-m \; t} F(t) - e^{- 2 m t} \left( e^{- k_+/2 \; t} G(t) + e^{k_+/2 \; t} H(t) \right)/(8 m k_1^2)}{2 m - e^{-2 m t} \left( e^{- k_+/2 \; t} G(t) + e^{k_+/2 \; t} H(t) \right)/(8 m k_1^2)},
\label{eq:SM_corr_toy}
\end{equation}
where $m = (\lambda+\eta)/2$ is the average rate, $k_+$ and $k_1$ are two real coefficients of the rates expressed below, and $F(t)$, $G(t)$ and $H(t)$ are the following functions:
\begin{equation*}
\begin{aligned}
&F(t) = 2m \cosh(k_2/2\;t) + (\lambda-\eta)^2 \sinh(k_2/2\;t)/k_2 \\
&G(t) = \left( 2m k_1 - k_+ (\lambda^2 + \eta^2) - 2 \lambda \eta k_-\right) \cos(k_-/2\;t) + \left( k_- (\lambda^2 + \eta^2) - 2 \lambda \eta k_+\right) \sin(k_-/2\;t) \\
&H(t) = \left( 2m k_1 + k_+ (\lambda^2 + \eta^2) + 2 \lambda \eta k_-\right) \cos(k_-/2\;t) + \left( k_- (\lambda^2 + \eta^2) - 2
\lambda \eta k_+\right) \sin(k_-/2\;t) \\
&k_1 = \sqrt{(\lambda-\eta)^4 + (4 \lambda \eta)^2} \\
&k_2 = \sqrt{(\lambda-\eta)^2-4 \lambda \eta} \\
&k_+ = \sqrt{\frac{k_1 + (\lambda-\eta)^2}{2}}\\
&k_- = \sqrt{\frac{k_1 - (\lambda-\eta)^2}{2}}.
\end{aligned}
\end{equation*}
Note that for some range of parameters $k_2$ can become imaginary.
In general, we will consider the regime of $\lambda \ll \eta$, which is implied by $N\mu \ll 1$, an inequality that makes $k_2$ a real number.

The behavior of this function is pretty complex, and both positive and negative correlations can be obtained depending on the parameters. Some examples are shown in Fig.~\ref{fig:SM_8states}b and \ref{fig:telegraph}.
Despite the function's complexity, we can make some observations about its behavior.
First, it is symmetric when switching $\lambda$ with $\eta$.
Second, we can inspect the behavior at extreme times.
For $t\rightarrow 0$ the correlation goes always to zero and also, the derivative in zero is always negative: $C(0)' = - \lambda \eta / (4 m)$. Therefore, a negative correlation is always expected for small time windows.
For long times the function goes to zero as well. We can state that the time scale that controls this collapse is the longest time scale among the exponential functions appearing in \ref{eq:SM_corr_toy}.
This time scale is $\tau_{max} = 1/(2m - k_2)$, such that $C(\Delta t) \rightarrow 0$ for $\Delta t \gg \tau_{max}$.
In the interesting limit in which the sweep regime holds, $N\mu \ll 1$, we have that $\lambda \ll \eta$.
The maximum time scale in this limit, at first order in $\lambda/\eta$ is $\tau_{max} \simeq 1/2 \lambda = T/2$.

\end{document}